\documentclass{article}

\usepackage[english]{babel}
\usepackage[utf8]{inputenc}
\usepackage{amsthm}

\usepackage{graphicx}
\usepackage{latexsym}
\usepackage{textcomp}
\usepackage{longtable}
\usepackage{multirow, booktabs}
\usepackage{natbib}
 \bibpunct{(}{)}{;}{a}{,}{,}
\usepackage{url}
\usepackage{hyperref}
\usepackage{amsmath, amsfonts, amssymb}
\usepackage{mathtools}
\usepackage{chemarrow}
\usepackage{subfigure}
\usepackage{rotating}
\usepackage{multirow}
\usepackage[ruled,vlined,noresetcount, boxed]{algorithm2e}
\usepackage{tikz}
\usepackage[margin=.75in]{geometry}

\newcommand{\etr}{\text{etr}}

\newcommand{\Exp}[1]{{\text{E}}[ \ensuremath{ #1 } ]  }

\newtheorem{theorem}{Theorem}
\newtheorem{theorem_appendix}{Theorem}
\newtheorem{corollary}{Corollary}
\newtheorem{example}{Example}
\newtheorem{example_appendix}{Example}

\theoremstyle{definition}
\newtheorem{definition}{Definition}[]

\title{Reducing Subspace Models for Large-Scale Covariance Regression}
\author{Alexander M. Franks}

\begin{document}
\maketitle
\begin{abstract}

We develop an envelope model for joint mean and covariance regression in the large $p$, small $n$ setting.  In contrast to existing envelope methods, which improve mean estimates by incorporating estimates of the covariance structure, we focus on identifying covariance heterogeneity by incorporating information about mean-level differences.  We use a Monte Carlo EM algorithm to identify a low-dimensional
subspace which explains differences in both means and covariances as a function of covariates, and then use MCMC to estimate the posterior uncertainty conditional on the inferred low-dimensional subspace. We demonstrate the utility of our model on a motivating application on the metabolomics of aging.  We also provide R code which can be used to develop and test other generalizations of the response envelope model.

\vfill
\end{abstract}

\section{Introduction}



Multivariate regression analyses are typically focused on how population means change with covariates.  However, there has been a growing need to infer covariance heterogeneity across observations as well.  For instance, in many applications mean-level effects may be small relative to subject variability.  In these settings, distributional differences may be more apparent from feature covariances.  Even when mean-level differences are large, better covariance estimates may lead to an improved understanding of the mechanisms underlying these apparent differences.  In this context, covariance regression models, in which the goal is to infer $\text{Cov}[Y\mid X]$, can be an important complement to existing regression analyses. In this paper, we propose a novel approach for joint mean and covariance regression in the setting in which the number of measured features may be larger than the number of observations (``large $p$, small $n$'').  Our  proposed method is motivated by an application in the metabolomics of aging \citep{kristal2005metabolomics}.  Metabolomics is a useful way to study age and age-related disease because the small molecules measured in metabolomic experiments represent the products of metabolism and reflect a detailed snapshot of physiological state of an organism.  Although there are many studies exploring how mean metabolite levels change with age, relatively little is understood about how the co-dependency of features changes with age \citep{le2019age}.

We gain traction on the large $p$, small $n$ problem by assuming that differences in high-dimensional outcomes are confined to a low-dimensional subspace.  Such an assumption is often well-motivated, especially in biological applications, where the measured molecules can be highly correlated due to their roles in a much smaller set of functional pathways \citep{liland2011multivariate, Heimberg2016}.  We formalize this idea by leveraging recent developments in the response envelope model \citep{cook2018principal}.  The original envelope model was developed in the context of efficient multivariate regression, where the goal is to recover the matrix of regression coefficients $\beta$ from an matrix of outcomes $Y$ given covariates $X$:
\begin{equation}
\label{eqn:lin_reg}
Y_{n\times p}=X_{n \times q}\beta_{q \times p} +\epsilon_{n \times p}
\end{equation}
\noindent typically with $\epsilon \sim N_p(0, \Sigma)$.  Even when $\Sigma$ is non-diagonal, the maximum likelihood solution to the multivariate regression is the same as the OLS solution \citep{Mardia1980}. As an alternative efficient strategy, \citet{cook2010envelope} proposed ``response envelopes'' for estimating $\beta$.  They posit a parametric link between the regression coefficients and the residual covariance:
\begin{align}
\label{eqn:env}
Y&=X\eta_{q \times s} V_{s\times p} + \epsilon\\ 
\Sigma&=V \Psi_1 V^{T}+V_{\perp} \Psi_{0} V_{\perp}^{T} \label{eqn:reducing}
\end{align}
where $\beta=\eta V$. $V_{p \times s}$ and $V_{\perp_{p \times (p-s)}}$ are orthogonal bases for complementary subspaces of dimension $s$ and $p-s$ respectively, with $V^TV_\perp=0$. The space spanned by $V$ corresponds to what Cook calls the ``material part'' because this subspace is relevant for inference on $\beta$. In contrast, the subspace spanned by $V_\perp$ is ``immaterial'' since $YV_\perp$ is invariant to $X$. Equation \ref{eqn:reducing} implies $V$ spans a \emph{reducing subspace} of $\Sigma$, since $\Sigma = VV^T\Sigma VV^T + V_\perp V_\perp^T\Sigma V_\perp V_\perp^T$  \citep{conway1990course}. When $V$ spans the smallest reducing subspace also satisfying Equation $\ref{eqn:env}$ it is called the $\Sigma$-envelope of $\beta$, denoted $\mathcal{E}_\Sigma(\beta)$, and reflects the subspace of material variation.  Most importantly, large efficiency gains for estimates of $\beta$  are possible over the OLS estimator when the dimension of $\mathcal{E}_\Sigma(\beta)$ is much smaller than $p$ and the leading eigenvalues of $\Psi_1$ are smaller than the leading eigenvalues of $\Psi_0$.  Variants and extensions of the classical response envelope model are well-summarized in recent review papers  \citep{lee2019review, cook2018principal}.




In these envelope-based methods, the focus is on efficient mean-level estimation.  While differences between multivariate means are useful for prediction and classification, our focus is on the covariance structure.  Different commonly used methods for covariance estimation provide unique insights about the underlying signal.  For example, in Gaussian graphical models, sparsity in the inverse covariance matrix captures conditional independence relationships  between variables \citep{Friedman2008, meinshausen2006}. \citet{danaher2014joint} extend the graphical models framework to jointly infer multiple sparse precision matrices with similar sparsity patterns. Graphical models are especially appealing when the system being analyzed is well-suited to graphical descriptions (e.g, as with biological pathways or social networks), but are inaccurate when important variables in the network are unobserved \citep{chandrasekaran2010latent}. 

In settings in which several relevant features are unobserved, factor models may be more appropriate.    Many variants of factor or principal components models have been proposed for characterizing covariance heterogeneity, especially in cases where the observations can be split across a finite set of discrete groups \citep{Flury1987, Schott1991, Boik2002, Hoff2009}.  More recent proposals incorporate reducing subspace assumptions by assuming that all differences between covariance matrices  are confined to a low-dimensional common subspace.  \citet{cook2008covariance}  proposed an early version of such covariance reducing models and \citet{franks2019shared} proposed an empirical Bayes generalization of their approach for the large $p$, small $n$ setting.  \citet{Wang2019} propose an even more general covariance reducing model for characterizing network heterogeneity via sparse inference of precision matrices,  whereas \citet{su2013estimation} consider covariance heterogeneity across groups in the response envelope model but still focus on inference for the mean in the $n > p$ setting.

When observations cannot be grouped according to unordered categorical variables more sophisticated methods are needed.  For example, in our motivating application, we expect feature co-dependency to change continuously with age.  Covariance regression methods are applicable in this setting, where the goal is to infer a mapping from q-dimensional predictor space to the space of $p \times p$ symmetric positive definite matrices.  \citet{Hoff2011} propose a linear covariance regression model to elucidate how covariability evolves jointly with the mean as a function of observed covariates and demonstrate the applicability of their approach with a 4-dimensional health outcome \citep{niu2019joint}.  \citet{fox2015bayesian} propose a non-linear generalization of the linear covariance regression.   However, these existing methods do not scale to high dimensional data and do not account for the potential efficiency gains from incorporating low-dimensional reducing subspace.  In this work, we bridge the gap between existing covariance reducing models and covariance regression models.  To this end, we propose a general envelope-based framework for exploring how large-scale covariance matrices vary with (possibly continuous) covariates in the large $p$, small $n$ setting.

\subsection{Contributions and Overview}

We generalize the shared subspace covariance model of \citep{franks2019shared} in two ways: 1) we explicitly incorporate a mean model and demonstrate empirically that including a mean can improve inference for covariance matrices and 2) we allow for covariance matrices to vary continuously with covariates, as opposed to discrete groups. We synthesize ideas from existing envelope and reducing subspace models \citep{lee2019review, cook2018principal, Wang2019} and covariance regression approaches \citep{Hoff2011, fox2015bayesian} and show that existing methods can be expressed as special cases of our more general framework. Our approach is applicable to high-dimensional settings without requiring sparsity in the observed features \citep[e.g. as in][]{su2016sparse}. In Section \ref{sec:model} we introduce relevant notation and describe the class of envelope models we use for covariance regression.   In Section \ref{sec:inference} we propose a maximum marginal likelihood approach for inferring envelopes, and show that objective functions used previously in the literature can be easily derived in this framework. We propose a Monte Carlo EM algorithm for the most general models for which no analytic objective function can be derived and provide R code which can be applied with any Bayesian model for the distribution of the projected data \citep{Franks_envelopeR_Git}.  In Section \ref{sec:simulation} we demonstrate the effectiveness of our framework in simulation, with a particular focus on the conditions under which mean-level estimates can in turn improve inference on covariance heterogeneity.  In Section \ref{sec:metabolomics}, we demonstrate the utility of our approach in our motivating application on a large $p$, small $n$ dataset of metabolite abundances in cerebrospinal fluid samples from nearly one hundred human subjects. We demonstrate how our model can be used to infer how correlations in metabolite abundances evolve with age and sex, and characterize functional metabolic groups that are associated with these changes.    

\section{Envelope Models for Joint Mean and Covariance Regression}
\label{sec:model}

\begin{figure}[t]
    \centering
    \subfigure[Projection in $\mathbb{R}^3$]{
    \label{fig:3dplot}
    \includegraphics[width=0.25\textwidth]{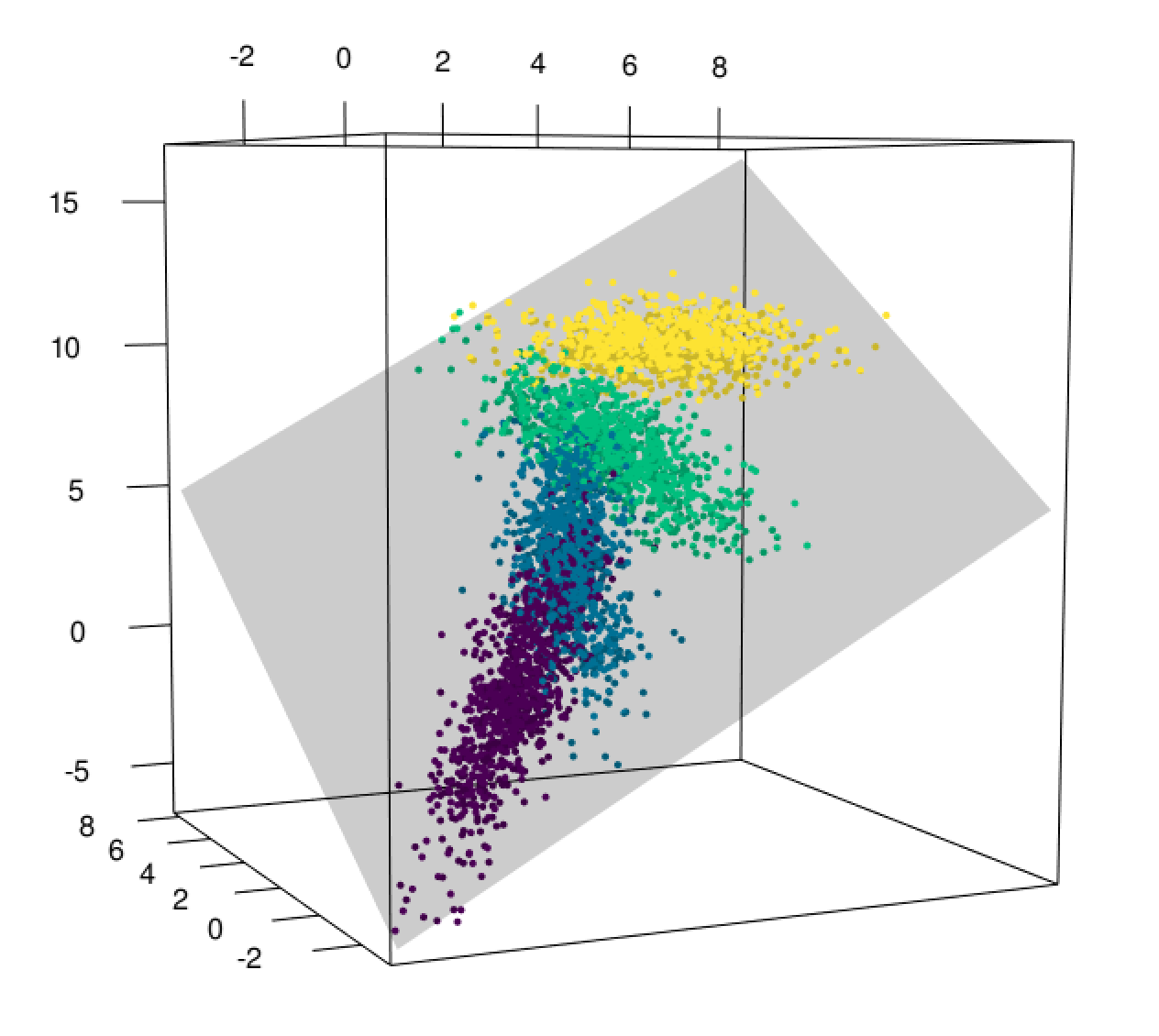}}
\quad
    \subfigure[$Y_xV$]{
    \label{fig:2dscatter}
    \includegraphics[width=0.2\textwidth]{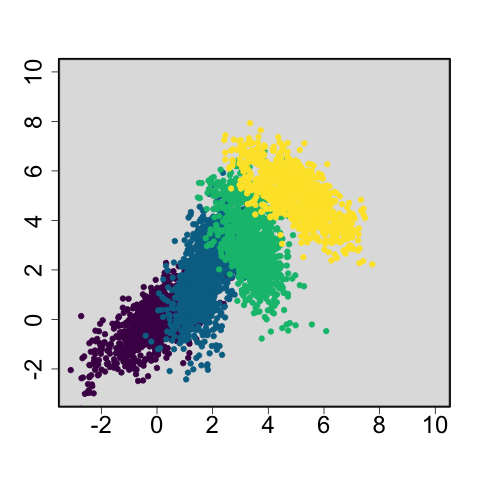}}
\quad 
\subfigure[$Y_xV_{\perp}$]{
\label{fig:2dscatterorth}
\includegraphics[width=0.2\textwidth]{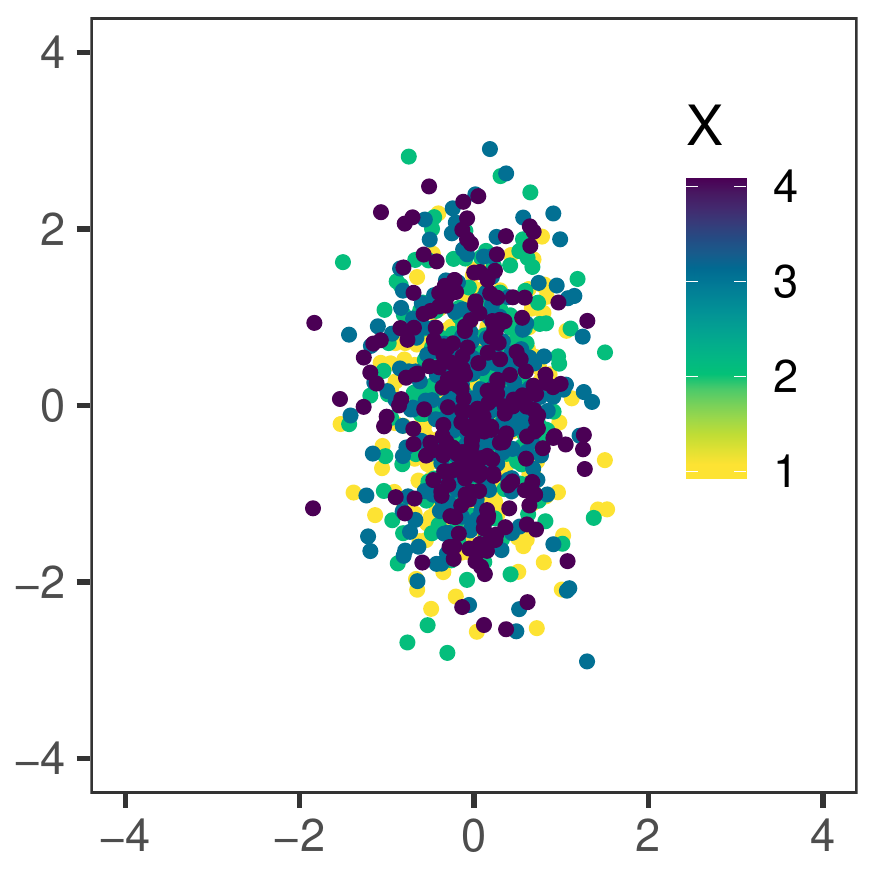}}
\caption{ An illustration of high-dimensional data, $Y$, colored according to discretized levels of a covariate, $X$, and projected into low-dimension subspaces.  Colors a) The projection of $Y$ into $\mathbb{R}^3$.  In this example, the
      differences in both the means and covariances of $Y_x$ are confined to a
      two-dimensional envelope, $\mathcal{E}_\mathcal{M}(\mathcal{U})$ (grey plane), spanned by $V$.  b) The
      data projected onto the $\mathcal{E}_\mathcal{M}(\mathcal{U})$,
      $Y_xV$, have means, $\phi_x$, and covariances $\Psi_x$ that co-vary with $x$. c) The orthogonal projection of the data, $Y_xV_{\perp}$
      has equal means and covariances across all $x$.  }
\label{fig:shared}
\end{figure}

Suppose that $Y$ is an $n \times p$ matrix of $n$ independent observations on $p$ correlated normally distributed features conditional on an $n \times q$ matrix of covariates,  $X$.  We use $y_i$ and $x_i$ to denote the $i$th row of $Y$ and $X$ respectively and allow both the the mean and covariance of $y_i$ to depend on $x_i$.  Here, $y_i \sim N(\alpha_0 + \mu_{x_i}, \Sigma_{x_i})$ and thus the joint density of $Y$ given $X$ is given by
\begin{equation}
p(Y | X, \Sigma, n) \propto l(\Sigma_x, \mu_x: Y, X)= \prod_i |\Sigma_{x_i}|^{-1 / 2} \operatorname{etr}\left(-\Sigma_x^{-1} (y_i - \alpha_0 - \mu_{x_i})^T(y_i-\alpha_0 - \mu_{x_i}) / 2\right)
\label{eqn:likelihood}
\end{equation}
\noindent where etr denotes the exponentiated trace.  Following the strategies proposed in envelope and shared subspace models, we make the assumption that there exists a low-dimensional subspace which includes all of the variation in $Y$ that is associated with $X$.  The projection of $Y$ onto the orthogonal subspace is invariant to $X$.  We        formalize this idea by making use of the following definition:

\begin{definition}
 Let $\mathcal{M} = \{\Sigma_{x_i} : i = 1, \ldots, n\}$ be the collection of $p \times p$ symmetric positive definite matrices matrices associated with the residual covariance of each observation $i$, and let $\mathcal{U} = \text{span}(\mu_{x_1}, \ldots, \mu_{x_n})$ be the span of the conditional means for each observation. We say that the $\mathcal{M}$-envelope of $\mathcal{U}$, $\mathcal{E}_{\mathcal{M}}({\mathcal{U}})$, is the intersection of all subspaces that contain $\mathcal{\mathcal{U}}$ and that reduce each member of $\mathcal{M}$. 
  \end{definition}
 
\noindent This is closely related to the definition proposed in \citet{su2013estimation}, with the difference that we do not presuppose a linear mean model for $\mu_x$, and we further allow $\Sigma_x$ to differ for all observations. To instantiate this idea in a model, we posit that:
\begin{align}
\label{eqn:shared_subspace}
 \mu_x &=  \phi_xV^T\\
\nonumber \Sigma_x &= V\Psi_xV^T + V_\perp\Psi_0V_\perp^T.
\end{align}

\noindent $V \in \mathcal{V}_{s, p}$, is an element of the Stiefel manifold, which consists of all $p \times s$ semi-orthogonal matrices and $V_\perp \in \mathcal{V}_{p-s, p}$ is a $p \times (p-s)$ semi-orthogonal matrix whose columns form the basis for the subspace of immaterial variation.  If $s$ is the smallest dimension satisfying Equation \ref{eqn:shared_subspace}, then the span of $V$ is the $\mathcal{M}$-envelope of $\mathcal{U}$,  the smallest subspace of material variation.  This is evident since $YV \sim N(\phi_x, \Psi_x \otimes I_n)$ depends on $x$, but $YV_\perp \sim N(0, \Psi_0 \otimes I_n)$ is invariant to changes in $X$ (see Figure \ref{fig:shared}).  $\phi_x$ is the $s$-dimensional projected-data mean, $E[YV \mid X=x]$ and $\Psi_x$ is the $s\times s$ projected-data covariance matrix, $\rm{Cov}(YV \mid X=x)$.  Throughout, we let model \ref{eqn:shared_subspace} be parameterized by $\theta$, where $\phi_x = f^\mu_\theta(x)$ is a function from $\mathbb{R}^q \to \mathbb{R}^{s}$ and $\Psi_x = f^\Sigma_\theta(x)$ a function from $\mathbb{R}^q \to \mathcal{S}^+_s$, the space of $s$-dimensional symmetric positive definite matrices.

Model \ref{eqn:shared_subspace} generalizes previously proposed response envelope models.  In the classic response envelope model, $f^\mu_\theta(x)$ is linear in $x$, i.e. $\phi_x = x\eta$ and $f_\theta^\Sigma(x)$ is constant i.e. $\Psi_x = \Psi_1$ \citep{cook2010envelope}.   \citet{su2013estimation} extend the response envelope model to allow $\Psi_x$ to vary with a categorical predictor in the $p \ll n$ setting, i.e. $\Psi_x = \Psi_k$ if $x=k$, for $k \in \{1, \ldots, K\}$.  \citet{franks2019shared} also assume categorical covariates, but focus on the multi-group covariance estimation problem in the $p > n$ setting with $f^\mu_\theta(x) = 0~ \forall x$. Due to the high dimensionality of problems considered, they take $\Psi_0$ to be diagonal so that $\Sigma_x$ follows the spiked covariance model \citep{Johnstone2001}.  


In many settings, more sophisticated assumptions about the structure of $\Psi_x$ are warranted.  For example, if the groups are related, a hierarchical covariance model on $\Psi_x$ may be more appropriate \citep{bouriga2013estimation, Hoff2009}.  We focus on the more general setting in which $\Psi_x$ varies with continuous covariates.  
\citet{Hoff2011} propose a linear covariance regression model which specifies that the differences in $\Psi_{x_i}$ and $\Psi_{x_j}$ for any $i \neq j$ can be described by a rank $K$ matrix.  Others have generalized this method for more flexible non-parametric models on $\Psi_x$ \citep{fox2015bayesian}.  

In this work, we make use of both the spiked covariance assumption and the linear covariance regression model and posit that
\begin{align}
    \Sigma_x &= V\Psi_xV^T + \sigma^2I \label{eqn:spiked_model}\\
    \Psi_x &= \sum^K_k B_k xx^T B_k + A  \label{eqn:linear_cov}
\end{align}

\noindent where $x$ is a $q$-vector, $B_k$ is an $s \times q$ matrix and A is a $s \times s$ symmetric positive semi-definite matrix. The first equation corresponds to the spiked covariance assumption and the second equation is the linear covariance model proposed by Hoff and Niu (2012, 2019).  Although the spiked model is unlikely to hold in practice, it is a particularly useful approximation in large $p$, small $n$ settings, where there is not enough power to accurately estimate more than a small number of factors of the covariance matrix \citep{Gavish2014}.  We assume the linear mean model $\phi_x = x\eta$ but this can also be generalized.  To our knowledge, our approach is the first envelope model proposed for joint mean and covariance regression.  


%


\section{Inference}
\label{sec:inference}

If a basis for the subspace of material variation, $V$, were known a priori, then we could estimate $\phi_x$ and $\Psi_x$ directly from the projected data $YV$.  This would be much more efficient than estimating parameters from the full data $Y$, especially when $s$ is small relative to $p$.  In practice, $V$ is not known, and thus must also be inferred.   As such, a two-stage approach is commonly used, in which we first estimate the subspace of material variation, and then conditional on the inferred subspace, estimate $\phi_x$ and $\Psi_x$.  $V$ is only identifiable up to rotations, since we can always reparameterize Equation \ref{eqn:shared_subspace} so that $\tilde \phi_x = \phi_x R$, $\tilde \Psi_x = R^T \Psi_X R$, and $\tilde V = VR$ for any $s \times s$ rotation matrix $R$.  However, the projection matrix $VV^T \in \mathcal{G}_{p, s}$ is identifiable, where  $\mathcal{G}_{p, s}$ is the Grassmannian manifold which consists of all $s$-dimensional subspaces of $\mathbb{R}^p$ \citep{Chikuse2012}. Although only the projection matrix $VV^T$ is identifiable, in practice most algorithms for subspace inference use what \citet{cook2010envelope} call a ``coordinate-based'' approach, by maximizing an objective function parameterized by a basis $V$ over the Stiefel manifold. Many efficient algorithms for coordinate-based envelope inference have been proposed \citep{Cook2015, Cook2016}. \citet{khare2017bayesian} propose an appealing method for joint Bayesian inference for all projected data parameters and the subspace in the response envelope  model.  However, Bayesian inference for semi-orthogonal matrices is challenging and their method does not scale for large $p$, e.g $p > 50$.  

In the context of their shared subspace covariance model, \citet{franks2019shared} propose a hybrid empirical Bayes approach, in which they first infer $V$ using maximum marginal likelihood, and conditional on that subspace use Bayesian inference to infer the projected data covariance matrices. Building on this work, we propose a general strategy based on maximum marginal likelihood for $V$ by specifying prior distributions for the projected data parameters $\phi_x$ and $\Psi_x$:

$$ [\hat V, \hat V_\perp] = \underset{[V~ V_\perp] \in \mathcal{V}_{p, p}}{\text{argmax}} \int_\Theta \mathcal{L}(V, V_\perp, \theta; Y, X)p(\theta) d\theta  $$ 

\noindent where $\mathcal{L}$ is the ``complete data" likelihood and the integration is with respect to the appropriate measure (e.g. on the product space of $R^s$ and $\mathcal{S}^+$ for Example \ref{eg:response_envelope} below). Conditional on $\hat V$, an estimate of a basis for the material subspace of variation, we then conduct Bayesian inference for $\phi_x$ and $\Psi_x$ from the $s$-dimensional projected data by sampling from the posterior, $P(\phi, \Psi \mid Y \hat V, X)$.  The maximum marginal likelihood approach with conjugate prior distributions yields objective functions which are closely related to those used in previous work. Below we highlight some of these examples.


\begin{example}[\textbf{Response Envelopes, \citet{cook2010envelope}}]
Assume the response envelope model (Equation \ref{eqn:env}). The response envelop model is a special case of model \ref{eqn:shared_subspace} with $\phi_x = X\eta$ with  $\Psi_x = \Psi_1$ independent of $X$.  Further, assume a priori that $\Psi_i \sim \text{inverse-Wishart}(U_i, \nu_i)$, $i \in \{0, 1\}$ and that $\beta = \eta V$ has  the multivariate matrix normal prior $\beta \mid (\Psi_1, V) \sim MN(B_0, V^T\Psi_1V \otimes \Lambda_0)$. Then the marginal log-likelihood for $(V, V_\perp)$, after integrating out $\theta = (\Psi_1, \Psi_0, \eta)$ is  
  \begin{equation}
  \label{eqn:marginal_lik}
      \ell(V, V_\perp) \propto -(n + s + \nu_1 + 1 - q)/2\log|V^T(A + U_1)V| - (n-(p-s) + \nu_0 -1)/2\log|V_\perp^T(Y^TY + U_0)V_\perp|
\end{equation}
where 
      $A = (Y - XB_n)^T(Y- X B_n) + B_n^T\Lambda_0B_n$ and $B_n = (X^TX + \Lambda_0)^{-1}X^TY$.
\label{eg:response_envelope}
\end{example}

\proof{ See appendix.}

With the non-informative improper prior distributions $U_i = \mathbf{0}$ and $\Lambda_0 = \mathbf{0}$ and $n \gg p$, this objective function is approximately proportional to the  objective function originally proposed by \citet{cook2010envelope}: $\ell(V, V_\perp) = \log(|V^TAV|) + \log(V_\perp^TY^TYV_\perp)$ .  In small $n$ settings, the difference in between these two objectives may be nontrivial.  It should also be noted Equation \ref{eqn:marginal_lik} can be expressed as a function of $V$ only, since the subspace spanned by $V_\perp$ is identified by $V$ alone \footnote{\citet{cook2013envelopes} use the modified objective $\ell(V) \propto \log|V^T(A )V| + \log|V^T(Y^TY )^{-1}V|$, which asymptotically has the same minimizer as Equation \ref{eqn:marginal_lik} when the envelope model holds.}.  The maximum marginal likelihood perspective also provides a principled way of including additional regularization by specifying appropriate  prior parameters \citep[see e.g.][]{yu2010comments}. \citet{su2016sparse} proposed an extension of to the envelope model to high-dimensional settings by a assuming a sparse envelope and modifying the usual envelope objective is augmented with a group lasso penalty on $V$.  In our framework, such a penalty would equivalently be viewed as prior distribution over the space of semi-orthogonal matrices, $V$.




\begin{example}[\textbf{Shared Subspace Covariance Models, \citet{franks2019shared}}]
Assume the shared subspace model proposed by \citet{franks2019shared}, that is $\phi_x = 0$ and $x$ is a single categorical predictor.  Let $Y_k$ be the $n_k \times p$ matrix of observations for which $x$ has level $k$ and let $\Psi_x = \Psi_k$ denote the corresponding covariance matrix of $Y_k$.  They assume a spiked covariance model, for which the data projected onto the subspace orthogonal to $V$ is isotropic, i.e. for which $\Psi_0 = \sigma^2I$.  $\Psi_k \sim inverse-Wishart(U_k, \nu_k)$ and $\sigma^2 \sim \text{inverse-Gamma}(\alpha, \kappa)$. Then the marginal log-likelihood, after integrating out $\theta = (\Psi_1, ... \Psi_k, \sigma^2)$ is
\begin{align*}
  \ell(V)  & \propto \sum_{k=1}^K  \left[-(n_k/2 - \nu_1)/2\log |V^T (Y_k^TY_k + U_k) V| - (n_k(p - s) + \alpha)\log(\rm{tr}[(I - V V^T)Y_k^TY_k]/2 + \kappa)\right]
\end{align*}
\end{example}
\noindent \citet{franks2019shared} propose a method for inferring the shared subspace using the EM algorithm, although the marginal distribution can be derived analytically, as above.  

\subsection{A Monte Carlo EM algorithm for Envelope Covariance Regression}
\label{sec:em}


In this Section we propose a general algorithm for inference in models satisfying Equation \ref{eqn:shared_subspace}.   We focus in particular on models for which $\Psi_x$ varies smoothly with continuous covariates although the algorithm can be applied more broadly.  Following the strategy described in the previous section, a reasonable approach would be to specify a prior distribution for $\mu_x$ and $\Psi_x$ and analytically marginalize out these variables to determine an objective function for $V$.  Unfortunately, for many covariance regression models, analytic marginalization is not possible.  To address this challenge, we make use of the Expectation Maximization (EM) algorithm \citep{dempster1977maximum}.   As a first step, we derive objective functions for $V$, a basis for the envelope assuming the projected data parameters, $\phi_x$ and $\Psi_x$ were known.  The following results give us the relevant objective functions.

\begin{theorem}

Assume a model satisfying Equation \ref{eqn:shared_subspace}.  Let the $\Psi_0 \sim \text{inverse-Wishart}(U_0, \nu_0)$, and assume the $\Psi_{x_i}$ and mean $\phi_{x_i}$ are known.  The marginal log-likelihood for $[V~ V_\perp]$, integrating over, $\Psi_0$, with $\Psi_{x_i}$ and $\phi_{x_i}$ assumed known is:
\begin{align}
\ell(V, V_\perp; Y, X, \Psi_x, \phi_x) &=  
    -\frac{1}{2}\sum_i\left[ y_i V \Psi_{x_i}^{-1}V^Ty_i^T - 2\phi_{x_i}\Psi_{x_i}^{-1}V^Ty_i^T\right]\\
\nonumber &~~~ - \frac{(n+(p-s)+\nu_0+1)}{2}\log|V_\perp^T(Y^TY + U_0)V_\perp|
\end{align}
\end{theorem}

When $p$ is large, the above objective function may lead to high variance estimators, since $Y^TY$ is a poor estimator of the marginaly covariance of $Y$ in the large $p$, small $n$ setting \citep{dempster1969elements, Stein1975}.  One option for high-dimensional inference is to enforce sparsity by penalizing the bases, $V$, with many non-zero rows.  Another alternative, which is consistent with the above model is to use a strong informative prior for $\Psi_0$ by specifying a positive semidefinite matrix $U_0$ and an appropriately large value for  $\nu_0$.  A related approach for dealing with large $p$, small $n$ data, we make use of the spiked covariance model described in Section \ref{sec:model} , Equation \ref{eqn:spiked_model}. Our proposed objective function for the spiked model is given in the following corollary.

\begin{corollary}

Assume a model satisfying Equation \ref{eqn:spiked_model} and a prior for $\sigma^2 \sim \text{inverse-Gamma}(\alpha, \kappa)$.  The marginal log-likelihood for $V$, integrating over $\sigma^2$, assuming $\Psi_{x_i}$ and $\phi_{x_i}$ are known is:
\begin{align}
\ell(V; Y, X, \Psi_x, \phi_x) &=  
    -\frac{1}{2}\sum_i\left[ y_i V \Psi_{x_i}^{-1}V^Ty_i^T - 2\phi_{x_i}\Psi_{x_i}^{-1}V^Ty_i^T\right]\\
\nonumber &~~~ - \left(\frac{(n(p-s)}{2} - \alpha\right)\log(\frac{1}{2}||Y||_F^2 - \frac{1}{2}||YV||_F^2 + \kappa)
\end{align}
\end{corollary}



\noindent Of course $\Psi_{x_i}$ and $\phi_{x_i}$ are not known in practice.  As such, we use of the EM algorithm for $V$, under an appropriate prior for the  model parameters for the projected data mean and covariance regression. Let $M^{(t)}_{x_i} = E[\phi_{x_i}\Psi^{-1}_{x_i}\mid V_t, x_i]$ and $K^{(t)}_{x_i} = E[\Psi^{-1}_{x_i}\mid V_t, x_i]$, where the expectation is taken with respect to $p(\theta \mid V_t, Y, X)$. We then make use of Corollary 1, replacing projected data parameters with conditional posterior mean estimates for the M-step of the EM algorithm:
\begin{align}
\label{eqn:mstep}
\ell(V, V_\perp; M^{(t)}_{x}, K^{(t)}_{x}) &= -\frac{1}{2}\sum_i\left[ y_i V K^{(t)}_{x_i}V^Ty_i^T - 2M^{(t)}_{x_i}V^Ty_i^T\right]  \\
\nonumber &~~~ - \left(\frac{(n(p-s)}{2} - \alpha\right)\log(\frac{1}{2}||Y||_F^2 - \frac{1}{2}||YV||_F^2 + \kappa)
\end{align}
%
The steps involved in the MCEM approach are described Algorithm \ref{alg:em}.  Below, we provide more detail about the E- and M-steps.

\begin{algorithm}[h]
\SetAlgoLined
 Initialize $V_0 \in \mathcal V_{p,s}$\;
    
 \While{$||V_t - V_{t-1}||_F > \epsilon$}{
    \textbf{E-step:} \\
    \For{$i \gets 1$ \textbf{to} $n$}{
         $M^{(t)}_{x_i} \gets E[\phi_{x_i}\Psi_{x_i}^{-1}\mid V_{t-1}, Y]$\;
         $K^{(t)}_{x_i} \gets E[\Psi_{x_i}^{-1}\mid V_{t-1}, Y]$\; 
    }
    \textbf{M-step:}\\
    \quad $[V_{t}, V_{\perp t} ]\gets \underset{[V~ V_\perp] \in \mathcal V_{p,s+r}}{\text{argmax }} \ell(V, V_{\perp}; M^{(t)}_{x}, K^{(t)}_{x})$,  Equation \ref{eqn:mstep}\;
 }
\caption{EM Subspace Estimation Algorithm}
\label{alg:em}
\end{algorithm}
\hfill

\noindent \textbf{M-Step:} As shown in Algorithm \ref{alg:em}, the M-step of the EM algorithm requires optimizing an objective function over $\mathcal{V}_{p, s}$.  Following existing work, we use the optimization method proposed by \citet{Wen2013} and implemented in the the package \texttt{rstiefel} \citep{rstiefel} to find a basis for the subspace that minimizes the log-complete likelihood (Equation \ref{eqn:mstep}). This feasible search algorithm has complexity of order $O(ps^2 + s^3)$, and as such can be very fast when the dimension of the envelope is much smaller than $p$.  This is usually assumed to be the case for large $p$, small $n$ problems, since we typically require that $s < n \ll p$.  The matrix derivative, $\frac{d\ell}{dV}$ which is required for optimization is given in the Appendix. \hfill 

\noindent \textbf{(Monte Carlo) E-Step:} The E-step involves computing $M^{(t)}_{x_i} = E[\phi_{x_i}\Psi_{x_i}^{-1}\mid V_{t-1}, Y]$ and $K^{(t)}_{x_i} = E[\Psi_{x_i}^{-1}\mid V_{t-1}, Y]$. However, for  arbitrary models on $\phi_{x_i}$ and $\Psi_{x_i}$, the expectations $M^{(t)}_{x}$ and $K^{(t)}_{x}$ are also not analytically tractable. Here, we propose a Monte Carlo EM algorithm (MCEM) \citep{levine2001implementations}.  We approximate $M^{(t)}_{x}$ and $K^{(t)}_{x}$ with MCMC samples at each iteration of Algorithm \ref{alg:em}.  Although MCMC is computationally expensive, for our motivating application we assume dimension of the envelope is small.  As such, Bayesian inference for $\phi_{x_i}$ and $\Psi^{-1}_{x_i}$ can be approximated quickly.  Importantly, any tractable Bayesian models for the inferring the posterior distribution $p(\phi_x, \Psi_x \mid YV, X=x)$ can be used in this framework.

Many models are possible for $\phi_x$ and $\Psi_x$.  
In this work, we demonstrate the utility of our method for Bayesian covariance regression  \citep{Hoff2009, fox2015bayesian}.  We apply Monte Carlo EM algorithm with the the covariance regression model of \citet{Hoff2011} for it's simplicity, interpretability, and the availability of R package implementation \texttt{covreg} \citep{covreg}.  Although we focus on covariance regression in this work, our R code implements Algorithm 1 for any Bayesian model for $\phi_x$ and $\psi_x$ for, provided a user supplied function to compute appropriate posterior means \citep{Franks_envelopeR_Git}.  


\subsection{Rank Selection and Initialization}
\label{eqn:rank_init}

\textbf{Rank selection:} A major challenge in any low rank method is choosing the appropriate rank.  Several model selection criteria can be used to help facilitate this choice.  Common approaches, including AIC, BIC, likelihood ratio tests and cross-validation, have all been applied in similar envelope models \citep{cook2010envelope, cook2013envelopes, Cook2016, Hoff2011}.  Following existing work, as a fast and useful heuristic, we propose applying asymptotically optimal (in mean squared error) singular value threshold for low rank matrix recovery with noisy data in the large $p$, small $n$ setting \citep{Gavish2014} and used a reducing covariance model by \citet{franks2019shared}.  This rank estimation procedures is motivated under spiked covariance model and is a function of median singular value of the data matrix and the ratio of the features to the sample size, $p/n$. When all of the covariates are categorical, our approach is equivalent to the rank selection procedure in  \citet{franks2019shared}. In Section \ref{sec:simulation} we explore the implications of misspecifying the envelope dimension.  

\noindent \textbf{Initialization:} Since optimization over the Stiefel manifold is non-convex, choosing a good (i.e. $\sqrt{n}$-consistent) initial estimate of $V$ is important \citep{Cook2016}.  Note that 
\begin{align*}
    \text{Cov}(Y) &= \beta\text{Cov}(X)\beta^T + E[\Sigma_X] \\
    &= V\left(\eta\text{Cov}(X)\eta^T + E[\Psi_x]\right)V^T + V_\perp \Psi_0 V_\perp^T
\end{align*}
so that a subset of the right singular vectors of $Y$ consistently estimate $\mathcal{E}_\mathcal{M}(\beta)$.  As such, one approach is to initialize $V$ to the first $s$ right singular vectors of $Y$.  In our analyses, we take the first $q$-columns of $V$ to be a basis for of $\hat \beta_{\text{OLS}}$, of the OLS solution.  For the remaining $s-q$ initial values of $V$, we choose the first $s-q$ right singular vectors of the residual $Y - X\hat \beta_{OLS}$.  \citet{Cook2015} consider a sequential 1D algorithm, in which each column of $V$ is updated in a coordinate-wise fashion.  They find that this fast way to initialization $V$ and could also be useful in our setting.


\section{Simulation Studies}
\label{sec:simulation}

In the response envelope model, there can be drastic efficiency gains for the mean regression coefficients, in particular when the envelope has small dimension and $||\Psi_0|| \gg ||\Psi_1||$ \citep{cook2010envelope}.  In this work, we explore the factors that improve efficiency of estimators for $\Psi_{x}$, focusing in this in this Section on the behavior and robustness of the proposed inferential approach in simulation.   We use the following model throughout: 
\begin{align}
\label{eqn:sim_model}
Y &=  X \mathbf{\eta}V + \epsilon_x\\
\nonumber \Psi_x &= \sum_k^{K=q} \Gamma_k x_ix_i^T\Gamma_k^T + \sigma^2I\\
\nonumber \epsilon_x &\sim N(0, V\Psi_xV^T + \sigma^2V_\perp V_\perp^T)
\end{align}
where $X_{n \times q} \sim N(0, 1)$, $\eta_{q \times s} \sim N(0, \tau^2)$, $\Gamma_{s \times q} \sim N(0, 1)$ are matrices with i.i.d random entries.  We set $n=100$ and $s=4$ and evaluate covariance estimates using Stein's loss, $L_S( \Psi_{x} , \hat\Psi_x) = \text{tr}( \Psi_x^{-1} \hat
\Psi_x ) - \log |\Psi_x^{-1} \hat \Psi_x | - p$ \citep{dey1985estimation}.  The Bayes estimator for Stein's loss is the inverse of
the posterior mean of the precision matrix, $\Exp{ \Psi_{x_i}^{-1} | Y, X, V}^{-1}$ which we estimate using MCEM (Section \ref{sec:em}).  In the simulations below, we compare the true covariance for observation $i$,  $V^T\Sigma_{x_i}V = \Psi_{x_i}$ to the fitted values 
\begin{align}
\hat \Psi_x &= V^T\hat \Sigma_{x_i}V\\
\nonumber &= (V^T \hat V) \Exp{ \Psi_{x_i}^{-1} | Y, X, \hat V}^{-1} \hat (\hat V ^T V)  + \hat \sigma^2 V^T\hat V_\perp \hat V_\perp^TV 
\end{align}


\noindent Accurate estimates of $\Psi_x$ depend on the accuracy of both the subspace estimates, $\hat V$, as well as estimates of projected data covariance matrices $\hat \Psi_x$.  In this Section we explore the robustness of our method to misspecification of chosen envelope dimension and also evaluate the effect that mean level differences have on inference for the covariance regression.

\begin{figure}[t]
    \centering
    \subfigure[Effect of specified envelope dimension]{
    \label{fig:steins_risk}
    \includegraphics[width=0.35\textwidth]{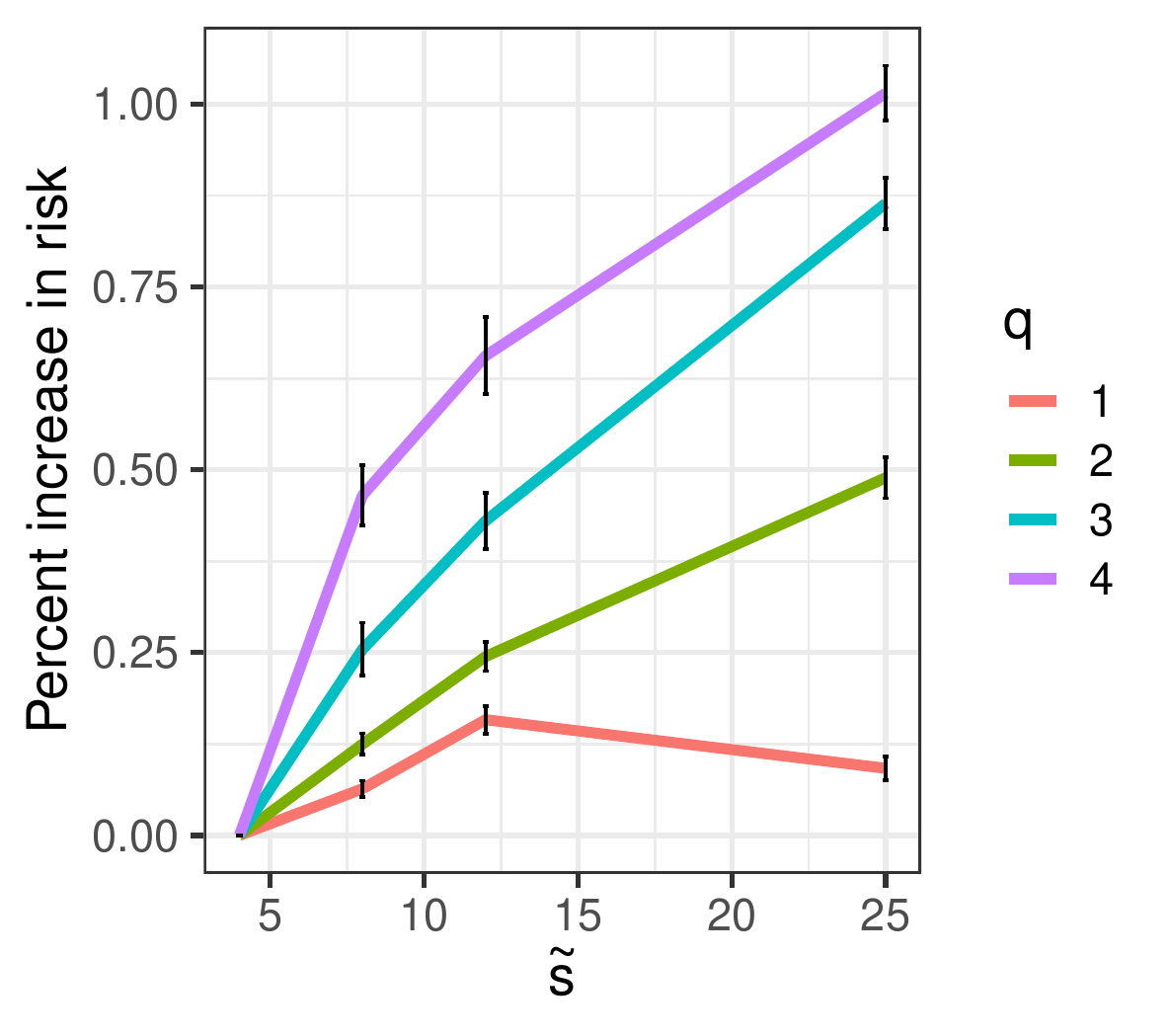}}
\quad
    \subfigure[Joint vs two-stage inference]{
    \label{fig:steins_risk2}
    \includegraphics[width=0.35\textwidth]{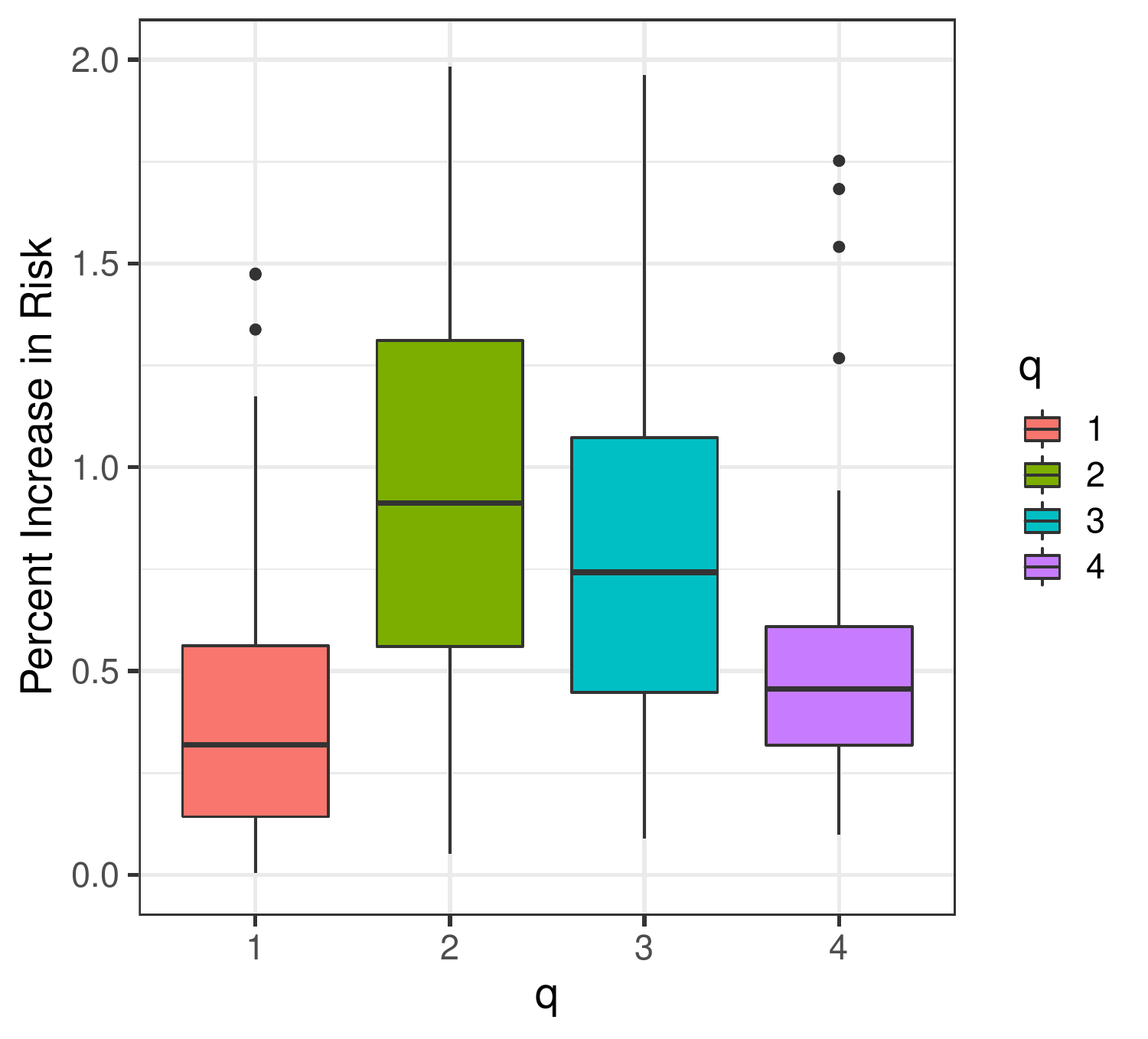}}
\caption{ a) Percentage increase in risk for covariance estimates due to misspecification of the subspace dimension relative to estimates using the true subspace dimension. We plot the mean percentage increase in loss as a function of the assumed subspace dimension, $\tilde s$ for data generated with $s=4$. Bars represent bootstrap 95\% confidence intervals.  b) Box plots of percentage increase in Stein's loss for covariance estimates as a function of the number of covariates.  When there are more covariates, mean-levels provide more information about the full subspace spanned by $V$.  In this simulation, this gain is partially offset by the increase in covariance parameters as a function of $q$.}
\label{fig:sim}
\end{figure}


\noindent \textbf{Misspecification of the Subspace Dimension:} \label{sec:misspec}  In practice, selecting the appropriate dimension, $s$, for the subspace of material variation is a challenging task.  In this simulation, we evaluate the increase in loss when fitting model with assumed envelope dimension  $\tilde s > s$.  Envelope covariance regression with $\tilde s = p$ is equivalent to running \texttt{covreg} on the full $p$-dimensional data.  Since inference in the covariance regression model is not computational tractable for large $\tilde s$, here we consider a relatively small-scale analysis: we generate data according to model proposed in Equation \ref{eqn:sim_model} where $s = 4$ and where $p=25$ but fit the model assuming $\tilde s \in \{4, 8, 12, 25\}$.  We set $\tau=3$ and $\sigma^2=1$ and calculate the average Stein's loss, $L_{\tilde s} = \frac{1}{n} \sum_i L_{\tilde s}( \Psi_{x_i} , \hat\Psi_{x_i})$ on $M=100$ i.i.d. datasets.  For each dataset, we fit our model at each value of $\tilde s$.

For the proposed data generating process, the distribution of Stein's loss is heavily right skewed.  In order to highlight the relative performance of the various estimators, we report the mean percentage increase in loss for all $\tilde s > s$, $\frac{1}{M}\sum_m^M \frac{L^m_{\tilde s} - L^m_{s}}{L^m{s}}$ relative to the loss for $\tilde s = s$, Figure \ref{fig:steins_risk}.  Although the true data generating model is nested in all models for which $\tilde s > s = 4$, there can be substantial efficiency gains when the chosen dimension is close to the true low dimensional subspace. The efficiency gains are also generally larger for larger values of $q$, since the mean regression provides more information about the envelope.  


\noindent \textbf{Quantifying the Effect of the Mean Differences on Covariance Estimation:} \label{sec:sim_mean}  Here, we further investigate the effect of the mean on covariance matrix inference.  We compare two approaches to inference: we consider a two stage approach in which we regress out the mean and then fit a covariance regression on the residual and compare this to the Monte Carlo EM algorithm for the envelope algorithm proposed in Section \ref{sec:inference}.  The two-stage approach ignores the parametric link between the mean and residual covariance and thus a loss of efficiency is expected.  We quantify the magnitude of this efficiency loss by again calculating Stein's loss  over $m=100$ i.i.d. datasets.  We run experiments with $s=4, q \in \{1, 2, 3, 4\}$ and $p=100$ and again plot percentage increase in Stein's loss for the two-stage approach relative to the MCEM envelope approach, Figure \ref{fig:steins_risk2}.  In general, $\text{rank}(\beta) \leq \text{rank}(V) = s$, with equality possible only if $q \geq s$. Our results show that efficient estimates of $\Psi_x$ are possible when $\text{rank}(\beta) \geq s$  and $||\beta||$ large, that is, when $\text{span}(V)$ can be identified precisely from mean levels differences alone. However, in the linear model, when $s \gg q$ the efficiency gains from incorporating information about mean differences may be more moderate.  In this simulation we also assume that $\Psi_x$ varies by a rank $q$ matrix so these gains are partially offset by the fact that there $q^2s$ parameters to infer for the covariance regression (Equation \ref{eqn:sim_model}).

\section{Application to metabolomic data}
\label{sec:metabolomics}

In this Section we demonstrate the utility of the subspace covariance regression model in our motivating application, an analysis of hetereogeneity in metabolomic data.  Metabolomics is the study of small molecules, known as metabolites, which include sugars, amino acids, and vitamins, amongst many others.  These small molecules represent the products of interactions between genes, proteins and the environment, and make up the structural and functional building blocks of all organisms.  As such, metabolomics can provide a detailed view into the physiological states of cells. By integrating measurements on the identities and abundances of these metabolites, we can develop a deeper understanding into the factors that shape phenotypic variation \citep{nicholson1999metabonomics, fiehn2002metabolomics, joyce2006model}.  Methods for analyzing metabolomic data stem from a long history of chemometrics, the field devoted to multivariate analysis of measurements on chemical systems \citep{wold1995chemometrics}.  There are many tools for multivariate analysis of metabolomics including partial least squares methods, metabolite set enrichment analyses, and methods based on graphical models \citep{sun2012covain, xia2015metaboanalyst}. Interestingly, partial least squares methodology, which has its origin in chemometrics, is closely related to envelope methods \citep{cook2013envelopes}.   

Although many tools have been developed to analyze large multivariate metabolomic datasets, the vast majority of these methods focus on quantifying changes in mean metabolite levels across conditions.  Quantifying metabolite co-dependence is a less studied problem, despite it's importance.  It is particularly relevant when there are unobserved factors which influence the relationship between covariates and the outcome. The effects of such unmeasured factors can manifest themselves as covariance heterogeneity.  Specifically, we can view the covariance regression model proposed by \citet{Hoff2011}  as a linear interaction model with unobserved covariates: their covariance regression model can be equivalently expressed as the linear model $Y = \mu_{x_i} + \sum_k B_k x_i z_{ik} + \epsilon$ where $x_i$ is an observed covariate (e.g. age) and $z_{ik} \sim N(0, 1)$ is an unobserved covariate (e.g. diet and lifestyle).

In this analysis, we demonstrate how estimates of covariance heterogeneity can improve our understanding of the biological mechanisms of aging.  There have been many metabolomic studies on aging and age-related diseases \citep{jove2014metabolomics, kristal2005metabolomics, kristal2007metabolomics}, most of which have focused on changes in mean-levels of metabolite.  However, even after accounting for variation in mean levels, significant heteoregeneity across chronological age groups remains  \citep{lowsky2014heterogeneity}, because unmeasured factors like diet, lifestyle and epigenetics  are thought to have different effects on the metabolome at different ages \citep{horvath2018dna, kristal2007metabolomics, brunet2017interaction, robinson2018determinants}.

Understanding how covariability amongst metabolites evolves with age is a largely unstudied problem, in part due to the challenges of modeling covariate-dependent covariance matrices.  One notable exception is \citet{le2019age}, who look at how age influences co-dependency among 50 metabolic biomarkers in large population of individuals ($n=27508$). They compute all pairwise Pearson correlations between biomarkers at yearly age-bins and analyze trends in these correlations.  They find that in general correlations tend to decrease with age. In this paper, we analyze the metabolomics of aging using our large $p$, small $n$ dataset using our envelope approach.   



\begin{figure}[t]
    \centering
    \subfigure[Residual metabolite correlations (increasing with age)]{
    \label{fig:pos_corrs}
    \includegraphics[width=0.45\textwidth]{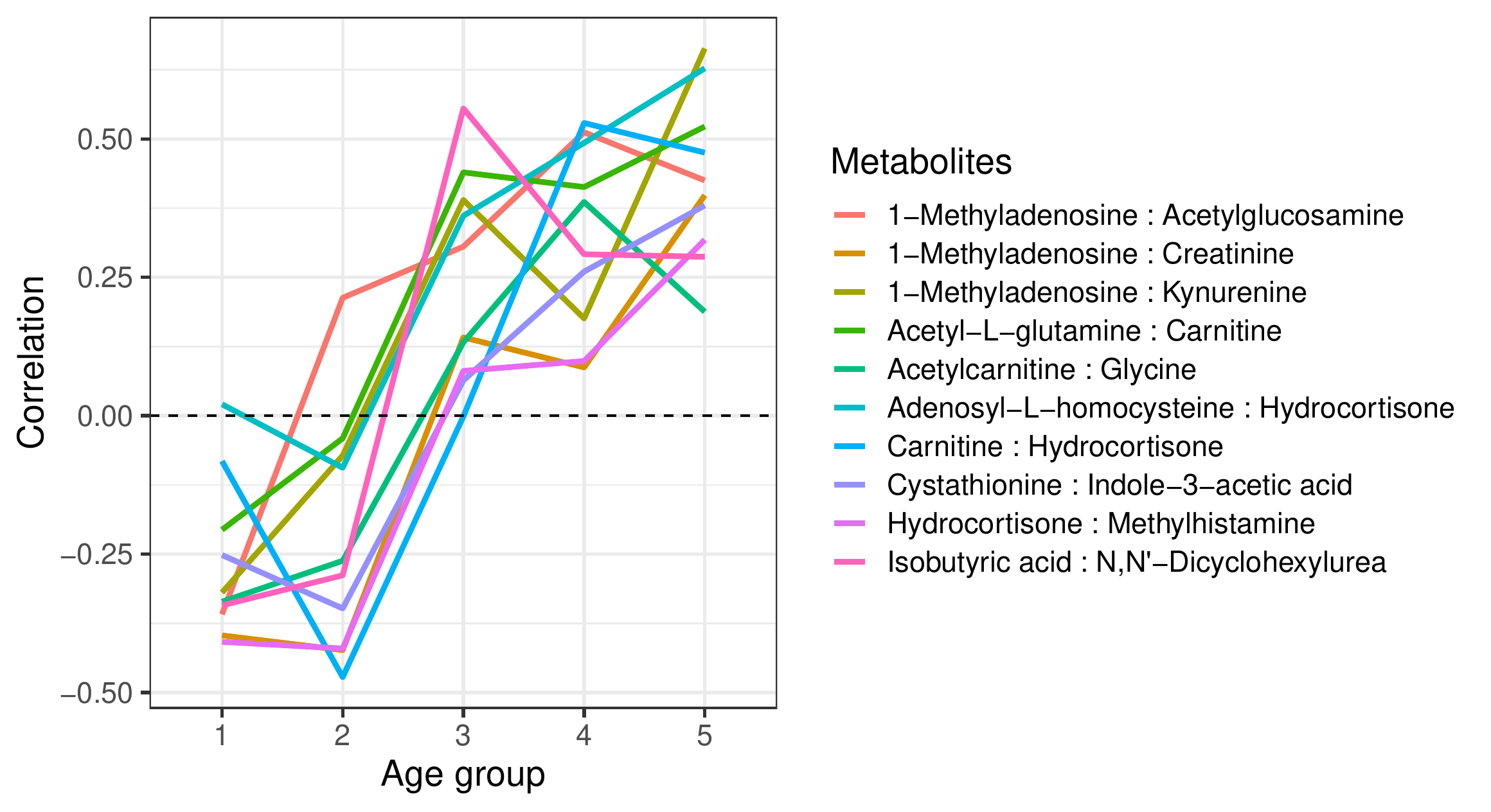}}
\quad
    \subfigure[Residual metabolite correlations (decreasing with age)]{
    \label{fig:neg_corrs}
    \includegraphics[width=0.45\textwidth]{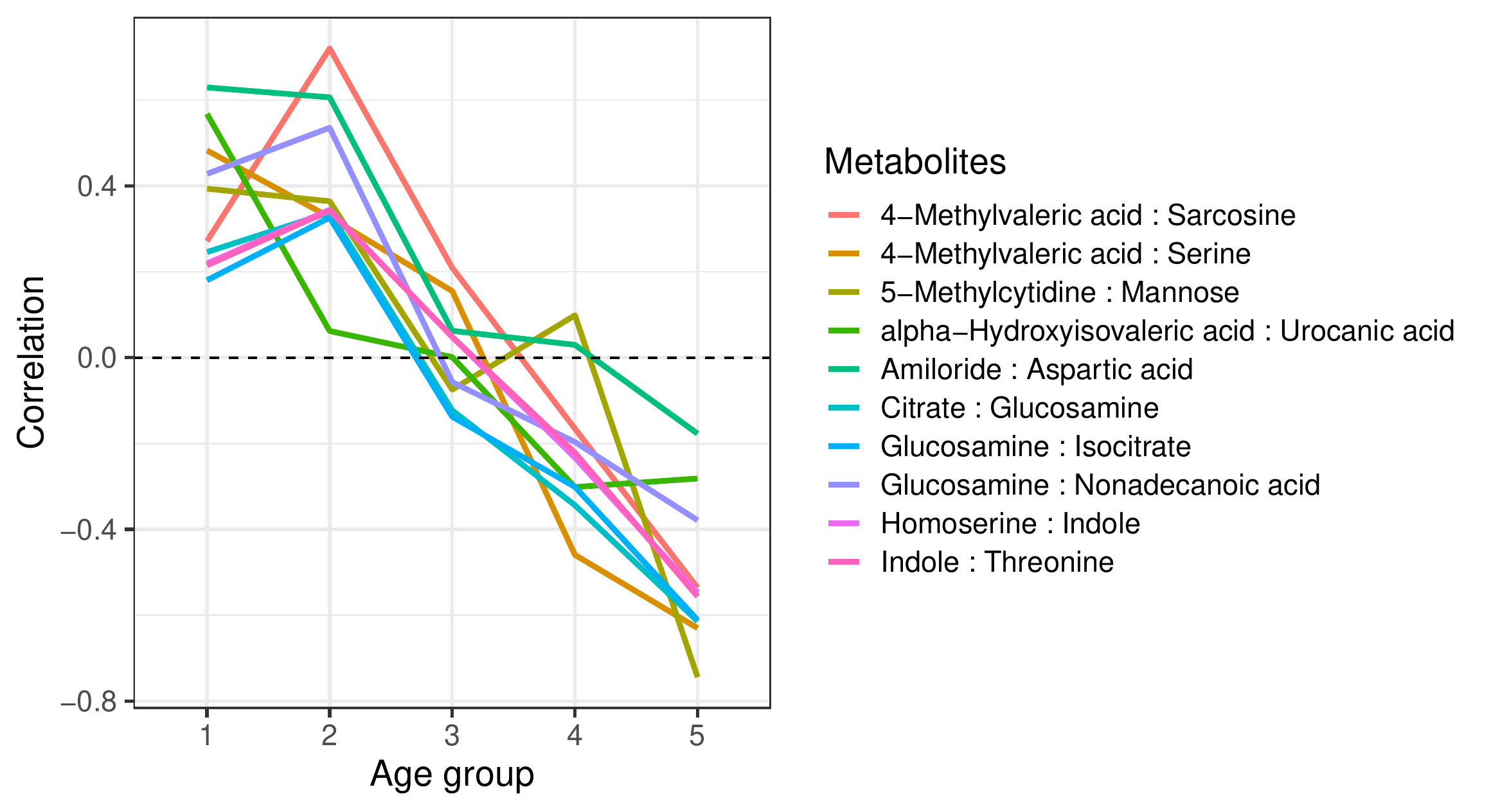}}
\caption{Residual Pearson correlations between metabolites show monotone trends after OLS regression. Observations are divided into five equal-sized groups, with age group 1 being the youngest individuals and age group 5 being the oldest individuals.  Some metabolite pairs indicate increasing Pearson correlation with age whereas others indicate decreasing correlations with age.}
\label{fig:corr_changes}
\end{figure}

Here, we analyze metabolomic assays on cerebrospinal fluid data from 85 human subjects at ages ranging from 20 years old to 86 years old.  The samples were assayed using both targeted and untargeted metabolite profiling LC-MS/MS experiments \citep{roberts2012targeted, dunn2013mass}. The targeted data includes 108 known metabolites and their corresponding abundances (i.e. $p \approx n$).  For the untargeted data we analyze approximately $2100$ metabolites which have measured abundances for at least 95\% of the subjects (i.e. $p \gg n$). For missing value imputation in the untargeted data, we use \texttt{Amelia}, a software package for imputation in multivariate normal data  \citep{Amelia}. \footnote{We leave it to future work to incorporate missing data mechanisms into our inferential algorithm.} In Figure \ref{fig:corr_changes} we depict exploratory plots which indicate that after OLS regression, pairwise correlations between residual metabolite abundances vary by age.  These plots suggest covariance regression is warranted with this data.
As such, we fit the envelope covariance regression model (Equations \ref{eqn:spiked_model} and \ref{eqn:linear_cov}) including both age and sex as covariates.  We use the rank-estimation method proposed by \citet{Gavish2014} and discussed in Section \ref{eqn:rank_init}.  For the targeted analysis we infer that $s=18$ and for the untargeted analysis $s=17$.

\begin{figure}
     \centering
     \begin{subfigure}
    \centering
    \includegraphics[width=0.75\textwidth]{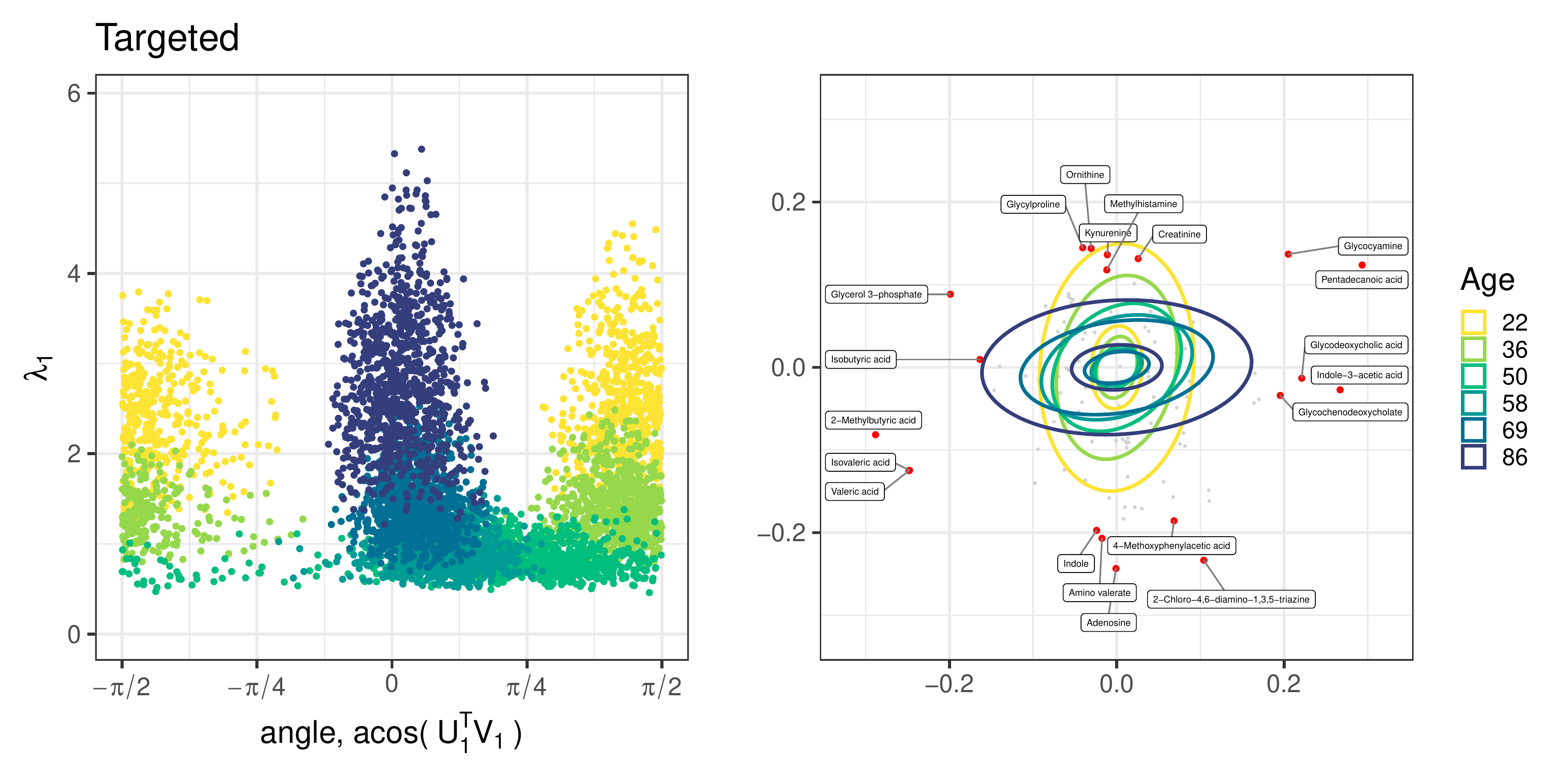}
\label{fig:aging_targeted}
     \end{subfigure} 
     \hfill \\
     \begin{subfigure}
         \centering
    \includegraphics[width=0.75\textwidth]{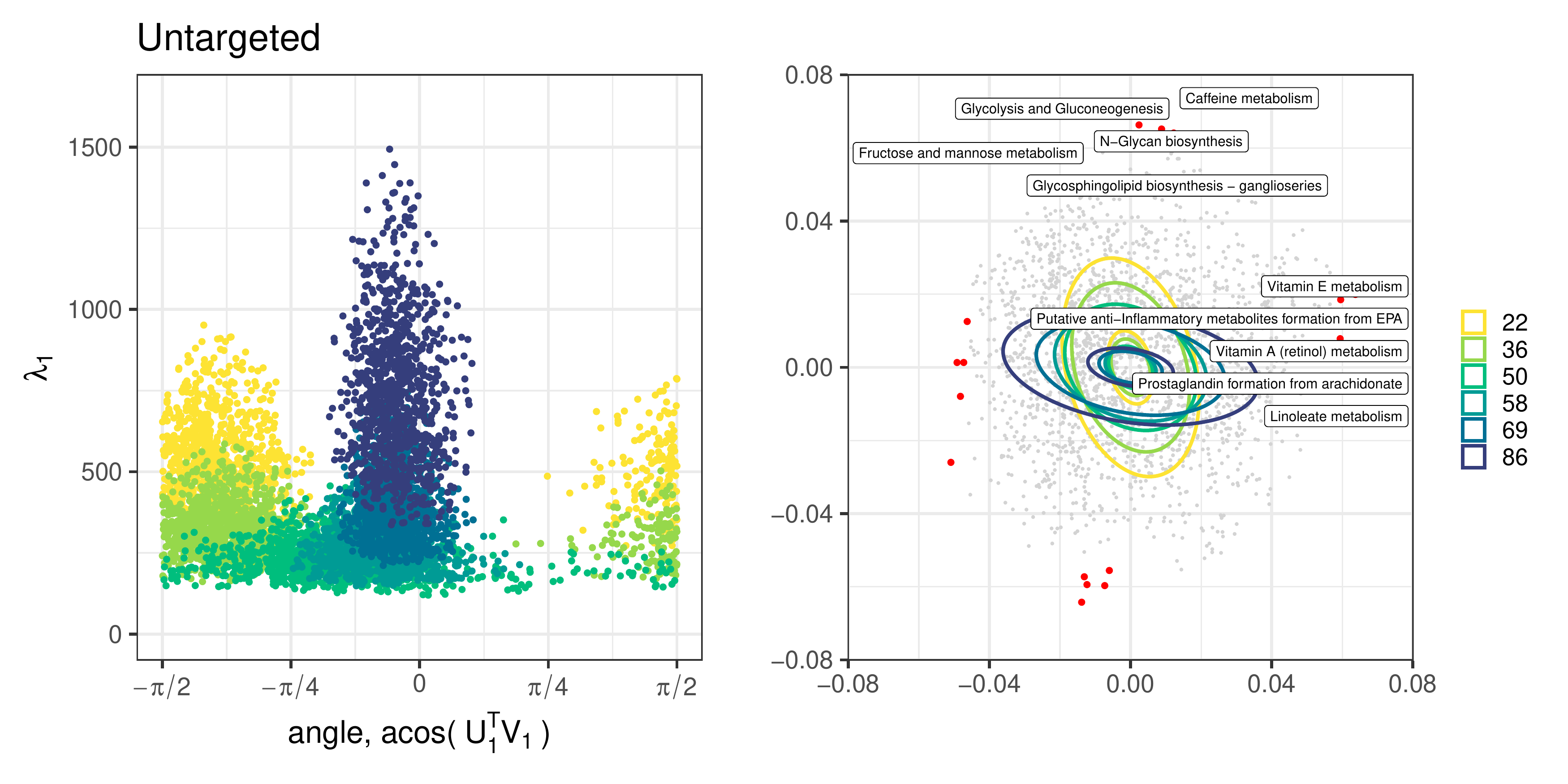}
\label{fig:aging_untargeted}
     \end{subfigure}
        \caption{Posterior summaries for age-dependent covariance matrices on the targeted (top) and untargeted (bottom) dataset.  Following \citet{franks2019shared}, we summarize the posterior distribution of the covariance matrices in terms of their eigenvalue and eigenvectors. Left) posterior samples of the largest eigenvalue and orientation of the first principal component on a two-dimensional subspace of $VV^T$ that describes the largest difference between the youngest and oldest male individuals. Right) A variant of the PCA biplot on the subspace of variation.  Contours depict the posterior mean covariance matrices at different ages.  Points reflect the metabolites with the largest factor loadings.  For targeted data we include the names of metabolites with the largest factor loadings, whereas for the untargeted analysis, we include pathways enrichment analysis \texttt{mummichog}.}
        \label{fig:results}
\end{figure}

Following an approach proposed by \citet{franks2019shared} we visualize posterior eigen-summaries on two dimensional subspaces of the envelope which are chosen to reflect the largest a posteriori significant differences between age groups.  Since the specific basis $\hat V$ for the inferred envelope is arbitrary, we make a change of basis to one whose first components reflect larger differences between chosen groups of interest.  For example, let $R$ denote the $s \times s$ orthogonal matrix of eigenvectors of the matrix $\hat \Psi_{\text{80,F}}-\hat \Psi_{\text{20,F}}$, the difference in projected ata covariance matrices for an 80 year old female and a 20 year old female.  Then let $\tilde V = \hat VR$ be the rotated basis for the inferred envelope, ordered according to directions which maximize the difference in the projected data covariance matrices of old (i.e. 80 years old) and young (i.e. 20 years old) females. In Figure \ref{fig:results} we summarize inferred covariance matrices by plotting covariance summary statistics $\tilde V^T \Sigma_x \tilde V$, for male subjects at six different ages at roughly equal intervals and include results for female samples in the Appendix. In the left column of Figure \ref{fig:results} we plot posterior samples of $\Psi_x$ colored by age.  Each point is a Monte Carlo sample of the largest eigenvalue $\lambda_1$ and the corresponding angle of the principal eigenvector, $U_1$, relative to the new coordinates defined by the first column of $\tilde V$.  The right column depicts a variant of a PCA biplot, with the contours illustrating the posterior mean of $\tilde V \hat \Psi_x \tilde V^T$.  The points indicate the feature loadings on two columns of $\tilde V$.  

The top panels of Figure \ref{fig:results} depict results from the targeted analysis ($p=108$) and the bottom panel depicts results from the untargeted analysis ($p=2113$).  In both of these figures, in older individuals there is higher variance among the metabolites with large loadings in the first component direction and lower variance for metabolites in the second direction.  The opposite is true for young individuals.  That the samples for the oldest and youngest individuals do not overlap suggests that these differences are significant \textit{a posteriori}.  For example, in the targeted analysis, we see that kynurenine and indole are anti-correlated along the second principal component direction, and have high variance in young individuals and lower variance in older individuals.  A related metabolite, indole-3-acetic acid has higher variance in older individuals.  These particular metabolites are associated with tryptophan metabolism, known to be one of the most important pathways associated with aging and age-related disease \citep{van2013tryptophan}.  Coefficients for these metabolites are also significant in the mean-regression (See Appendix).

For the untargeted data, metabolite identities are not known, only their mass-to-charge ratio and retention times.  As such, we use the package \texttt{mummichog} for inferring significantly enriched metabolomic functional groups from high throughput, untargeted metabolomic data \citep{li2013predicting}. We use the absolute value of the loadings along each principle axis (the columns of $\tilde V$) as inputs to \texttt{mummichog}, since the values represent correlated features with high variability.  Significantly enriched pathways associated with each principal axis are noted in the figure on the bottom right.  For older individuals, we infer that there is higher variance in metabolites associated with pathways listed on the right (e.g. Vitamin A or E metabolism).  In contrast, younger individuals seem to have more variable metabolite levels in pathways associated with the pathways listed on the top (e.g. Glycolysis and Gluconeogenesis or fructose and mannose metabolism).  We also ran pathway enrichment on the mean coefficients for both age and sex (see Appendix).  Some of the enriched pathways that were found to have metabolites with significant mean-level differences differ from those identified using the inferred covariance parameters.  This suggests that additional insight can be gained by analyzing covariance heterogeneity that cannot be gleaned from mean level differences alone.  

\section{Discussion}

In this paper, we extend the classic response envelope model to settings in which residual covariance matrices vary with continuous predictors.  Unlike previous work, which primarily focuses on mean estimation in the linear model, we demonstrate that envelope models can also be used to improve inferences for covariance heterogeneity in the large $p$, small $n$ setting.  Our work extends the hetereoskedastic envelope model \citep{su2013estimation} and shared subspace of model of \citet{franks2019shared} by accounting for continuously varying covariates.  We demonstrate our approach using a linear Bayesian covariance regression model  \citet{Hoff2011, niu2019joint}, but our method is compatible with any Bayesian model for joint mean and covariance regression.  We can also incorporate many of the extensions proposed in existing literature.  For example, sparsifying penalities can be used as an alternative to the spiked model for high-dimensional inference \citet{su2016sparse} and can easily be integrated into the objective function.  Although we assume normally distributed data, we can also extend the shared subspace approach to account for heavy tailed data, by modeling the outcome as a scale mixture of normal distributions \citep[e.g. see][]{ding2019envelope}.  

Finally, while we focus on linear mean and covariance models in this work, the Monte Carlo EM framework can be extended to non-linear and nonparametric models by leveraging other existing models for Bayesian inference in this setting.  
Tree-based models, like multivariate Bayesian Additive Regress Trees (BART) \citep{chipman2010bart} could be used as a non-parametric alternative for $\phi_x$.  As noted, we have implemented the Monte Carlo EM algorithm as a practical tool for inference in many of these more elaborate envelope-based models \citep{Franks_envelopeR_Git}.








\bibliographystyle{chicago}
\bibliography{refs.bib}

\begin{thebibliography}{}

\bibitem[\protect\citeauthoryear{Boik}{Boik}{2002}]{Boik2002}
Boik, R.~J. (2002).
\newblock Spectral models for covariance matrices.
\newblock {\em Biometrika\/}~{\em 89\/}(1), 159--182.

\bibitem[\protect\citeauthoryear{Bouriga and F{\'e}ron}{Bouriga and
  F{\'e}ron}{2013}]{bouriga2013estimation}
Bouriga, M. and O.~F{\'e}ron (2013).
\newblock Estimation of covariance matrices based on hierarchical
  inverse-wishart priors.
\newblock {\em Journal of Statistical Planning and Inference\/}~{\em 143\/}(4),
  795--808.

\bibitem[\protect\citeauthoryear{Brunet and Rando}{Brunet and
  Rando}{2017}]{brunet2017interaction}
Brunet, A. and T.~A. Rando (2017).
\newblock Interaction between epigenetic and metabolism in aging stem cells.
\newblock {\em Current opinion in cell biology\/}~{\em 45}, 1--7.

\bibitem[\protect\citeauthoryear{Chandrasekaran, Parrilo, and
  Willsky}{Chandrasekaran et~al.}{2010}]{chandrasekaran2010latent}
Chandrasekaran, V., P.~A. Parrilo, and A.~S. Willsky (2010).
\newblock Latent variable graphical model selection via convex optimization.
\newblock In {\em 2010 48th Annual Allerton Conference on Communication,
  Control, and Computing (Allerton)}, pp.\  1610--1613. IEEE.

\bibitem[\protect\citeauthoryear{Chikuse}{Chikuse}{2012}]{Chikuse2012}
Chikuse, Y. (2012).
\newblock {\em Statistics on special manifolds}, Volume 174.
\newblock Springer Science \& Business Media.

\bibitem[\protect\citeauthoryear{Chipman, George, McCulloch, et~al.}{Chipman
  et~al.}{2010}]{chipman2010bart}
Chipman, H.~A., E.~I. George, R.~E. McCulloch, et~al. (2010).
\newblock Bart: Bayesian additive regression trees.
\newblock {\em The Annals of Applied Statistics\/}~{\em 4\/}(1), 266--298.

\bibitem[\protect\citeauthoryear{Conway}{Conway}{1990}]{conway1990course}
Conway, J.~B. (1990).
\newblock A course in functional analysis. 1990.
\newblock {\em Graduate Texts in Mathematics\/}.

\bibitem[\protect\citeauthoryear{Cook, Helland, and Su}{Cook
  et~al.}{2013}]{cook2013envelopes}
Cook, R., I.~Helland, and Z.~Su (2013).
\newblock Envelopes and partial least squares regression.
\newblock {\em Journal of the Royal Statistical Society: Series B (Statistical
  Methodology)\/}~{\em 75\/}(5), 851--877.

\bibitem[\protect\citeauthoryear{Cook}{Cook}{2018}]{cook2018principal}
Cook, R.~D. (2018).
\newblock Principal components, sufficient dimension reduction, and envelopes.

\bibitem[\protect\citeauthoryear{Cook and Forzani}{Cook and
  Forzani}{2008}]{cook2008covariance}
Cook, R.~D. and L.~Forzani (2008).
\newblock Covariance reducing models: An alternative to spectral modelling of
  covariance matrices.
\newblock {\em Biometrika\/}~{\em 95\/}(4), 799--812.

\bibitem[\protect\citeauthoryear{Cook, Forzani, and Su}{Cook
  et~al.}{2016}]{Cook2016}
Cook, R.~D., L.~Forzani, and Z.~Su (2016).
\newblock {A note on fast envelope estimation}.
\newblock {\em Journal of Multivariate Analysis\/}~{\em 150}, 42--54.

\bibitem[\protect\citeauthoryear{Cook, Li, and Chiaromonte}{Cook
  et~al.}{2010}]{cook2010envelope}
Cook, R.~D., B.~Li, and F.~Chiaromonte (2010).
\newblock Envelope models for parsimonious and efficient multivariate linear
  regression.
\newblock {\em Statistica Sinica\/}, 927--960.

\bibitem[\protect\citeauthoryear{Cook and Zhang}{Cook and
  Zhang}{2015}]{Cook2015}
Cook, R.~D. and X.~Zhang (2015).
\newblock {Algorithms for Envelope Estimation}.
\newblock {\em Journal of Computational and Graphical Statistics\/}~{\em
  8600\/}(March), 00--00.

\bibitem[\protect\citeauthoryear{Danaher, Wang, and Witten}{Danaher
  et~al.}{2014}]{danaher2014joint}
Danaher, P., P.~Wang, and D.~M. Witten (2014).
\newblock The joint graphical lasso for inverse covariance estimation across
  multiple classes.
\newblock {\em Journal of the Royal Statistical Society: Series B (Statistical
  Methodology)\/}~{\em 76\/}(2), 373--397.

\bibitem[\protect\citeauthoryear{Dempster}{Dempster}{1969}]{dempster1969elements}
Dempster, A.~P. (1969).
\newblock Elements of continuous multivariate analysis.
\newblock Technical report.

\bibitem[\protect\citeauthoryear{Dempster, Laird, and Rubin}{Dempster
  et~al.}{1977}]{dempster1977maximum}
Dempster, A.~P., N.~M. Laird, and D.~B. Rubin (1977).
\newblock Maximum likelihood from incomplete data via the em algorithm.
\newblock {\em Journal of the Royal Statistical Society: Series B
  (Methodological)\/}~{\em 39\/}(1), 1--22.

\bibitem[\protect\citeauthoryear{Dey, Srinivasan, et~al.}{Dey
  et~al.}{1985}]{dey1985estimation}
Dey, D.~K., C.~Srinivasan, et~al. (1985).
\newblock Estimation of a covariance matrix under stein's loss.
\newblock {\em The Annals of Statistics\/}~{\em 13\/}(4), 1581--1591.

\bibitem[\protect\citeauthoryear{Ding, Su, Zhu, and Wang}{Ding
  et~al.}{2019}]{ding2019envelope}
Ding, S., Z.~Su, G.~Zhu, and L.~Wang (2019).
\newblock Envelope quantile regression.
\newblock {\em Statistica Sinica\/}.

\bibitem[\protect\citeauthoryear{Dunn, Erban, Weber, Creek, Brown, Breitling,
  Hankemeier, Goodacre, Neumann, Kopka, et~al.}{Dunn
  et~al.}{2013}]{dunn2013mass}
Dunn, W.~B., A.~Erban, R.~J. Weber, D.~J. Creek, M.~Brown, R.~Breitling,
  T.~Hankemeier, R.~Goodacre, S.~Neumann, J.~Kopka, et~al. (2013).
\newblock Mass appeal: metabolite identification in mass spectrometry-focused
  untargeted metabolomics.
\newblock {\em Metabolomics\/}~{\em 9\/}(1), 44--66.

\bibitem[\protect\citeauthoryear{Fiehn}{Fiehn}{2002}]{fiehn2002metabolomics}
Fiehn, O. (2002).
\newblock Metabolomics—the link between genotypes and phenotypes.
\newblock In {\em Functional genomics}, pp.\  155--171. Springer.

\bibitem[\protect\citeauthoryear{Flury}{Flury}{1987}]{Flury1987}
Flury, B. (1987).
\newblock {Two generalizations of the common principal component model}.
\newblock {\em Biometrika\/}~{\em 74\/}(1), 59--69.

\bibitem[\protect\citeauthoryear{Fox and Dunson}{Fox and
  Dunson}{2015}]{fox2015bayesian}
Fox, E.~B. and D.~B. Dunson (2015).
\newblock Bayesian nonparametric covariance regression.
\newblock {\em The Journal of Machine Learning Research\/}~{\em 16\/}(1),
  2501--2542.

\bibitem[\protect\citeauthoryear{Franks}{Franks}{2020}]{Franks_envelopeR_Git}
Franks, A.~M. (2020).
\newblock enveloper.
\newblock \url{https://github.com/afranks86/envelopeR}.

\bibitem[\protect\citeauthoryear{Franks and Hoff}{Franks and
  Hoff}{2019}]{franks2019shared}
Franks, A.~M. and P.~Hoff (2019).
\newblock Shared subspace models for multi-group covariance estimation.
\newblock {\em Journal of Machine Learning Research\/}~{\em 20\/}(171), 1--37.

\bibitem[\protect\citeauthoryear{Friedman, Hastie, and Tibshirani}{Friedman
  et~al.}{2008}]{Friedman2008}
Friedman, J., T.~Hastie, and R.~Tibshirani (2008).
\newblock {Sparse inverse covariance estimation with the graphical lasso}.
\newblock {\em Biostatistics\/}~{\em 9\/}(3), 432--441.

\bibitem[\protect\citeauthoryear{Gavish and Donoho}{Gavish and
  Donoho}{2014}]{Gavish2014}
Gavish, M. and D.~L. Donoho (2014).
\newblock The optimal hard threshold for singular values is {$4/\sqrt{3}$}.
\newblock {\em Information Theory, IEEE Transactions on\/}~{\em 60\/}(8),
  5040--5053.

\bibitem[\protect\citeauthoryear{Heimberg, Bhatnagar, El-Samad, and
  Thomson}{Heimberg et~al.}{2016}]{Heimberg2016}
Heimberg, G., R.~Bhatnagar, H.~El-Samad, and M.~Thomson (2016).
\newblock Low dimensionality in gene expression data enables the accurate
  extraction of transcriptional programs from shallow sequencing.
\newblock {\em Cell Systems\/}~{\em 2\/}(4), 239--250.

\bibitem[\protect\citeauthoryear{Hoff and Franks}{Hoff and
  Franks}{2019}]{rstiefel}
Hoff, P. and A.~Franks (2019).
\newblock {\em rstiefel: Random Orthonormal Matrix Generation and Optimization
  on the Stiefel Manifold}.
\newblock R package version 1.0.0.

\bibitem[\protect\citeauthoryear{Hoff}{Hoff}{2009}]{Hoff2009}
Hoff, P.~D. (2009).
\newblock A hierarchical eigenmodel for pooled covariance estimation.
\newblock {\em Journal of the Royal Statistical Society: Series B (Statistical
  Methodology)\/}~{\em 71\/}(5), 971--992.

\bibitem[\protect\citeauthoryear{Hoff and Niu}{Hoff and Niu}{2012}]{Hoff2011}
Hoff, P.~D. and X.~Niu (2012).
\newblock A covariance regression model.
\newblock {\em Statistica Sinica\/}~{\em 22}, 729--753.

\bibitem[\protect\citeauthoryear{Honaker, King, and Blackwell}{Honaker
  et~al.}{2011}]{Amelia}
Honaker, J., G.~King, and M.~Blackwell (2011).
\newblock {Amelia II}: A program for missing data.
\newblock {\em Journal of Statistical Software\/}~{\em 45\/}(7), 1--47.

\bibitem[\protect\citeauthoryear{Horvath and Raj}{Horvath and
  Raj}{2018}]{horvath2018dna}
Horvath, S. and K.~Raj (2018).
\newblock Dna methylation-based biomarkers and the epigenetic clock theory of
  ageing.
\newblock {\em Nature Reviews Genetics\/}~{\em 19\/}(6), 371.

\bibitem[\protect\citeauthoryear{Johnstone}{Johnstone}{2001}]{Johnstone2001}
Johnstone, I.~M. (2001).
\newblock On the distribution of the largest eigenvalue in principal components
  analysis.
\newblock {\em Annals of statistics\/}, 295--327.

\bibitem[\protect\citeauthoryear{Jov{\'e}, Portero-Ot{\'\i}n, Naud{\'\i},
  Ferrer, and Pamplona}{Jov{\'e} et~al.}{2014}]{jove2014metabolomics}
Jov{\'e}, M., M.~Portero-Ot{\'\i}n, A.~Naud{\'\i}, I.~Ferrer, and R.~Pamplona
  (2014).
\newblock Metabolomics of human brain aging and age-related neurodegenerative
  diseases.
\newblock {\em Journal of Neuropathology \& Experimental Neurology\/}~{\em
  73\/}(7), 640--657.

\bibitem[\protect\citeauthoryear{Joyce and Palsson}{Joyce and
  Palsson}{2006}]{joyce2006model}
Joyce, A.~R. and B.~{\O}. Palsson (2006).
\newblock The model organism as a system: integrating'omics' data sets.
\newblock {\em Nature reviews Molecular cell biology\/}~{\em 7\/}(3), 198--210.

\bibitem[\protect\citeauthoryear{Khare, Pal, Su, et~al.}{Khare
  et~al.}{2017}]{khare2017bayesian}
Khare, K., S.~Pal, Z.~Su, et~al. (2017).
\newblock A bayesian approach for envelope models.
\newblock {\em The Annals of Statistics\/}~{\em 45\/}(1), 196--222.

\bibitem[\protect\citeauthoryear{Kristal and Shurubor}{Kristal and
  Shurubor}{2005}]{kristal2005metabolomics}
Kristal, B.~S. and Y.~I. Shurubor (2005).
\newblock Metabolomics: opening another window into aging.
\newblock {\em Science of aging knowledge environment: SAGE KE\/}~{\em
  2005\/}(26), pe19--pe19.

\bibitem[\protect\citeauthoryear{Kristal, Shurubor, Kaddurah-Daouk, and
  Matson}{Kristal et~al.}{2007}]{kristal2007metabolomics}
Kristal, B.~S., Y.~I. Shurubor, R.~Kaddurah-Daouk, and W.~R. Matson (2007).
\newblock Metabolomics in the study of aging and caloric restriction.
\newblock In {\em Biological Aging}, pp.\  393--409. Springer.

\bibitem[\protect\citeauthoryear{Le~Goallec and Patel}{Le~Goallec and
  Patel}{2019}]{le2019age}
Le~Goallec, A. and C.~J. Patel (2019).
\newblock Age-dependent co-dependency structure of biomarkers in the general
  population of the united states.
\newblock {\em Aging (Albany NY)\/}~{\em 11\/}(5), 1404.

\bibitem[\protect\citeauthoryear{Lee and Su}{Lee and Su}{2019}]{lee2019review}
Lee, M. and Z.~Su (2019).
\newblock A review of envelope models.

\bibitem[\protect\citeauthoryear{Levine and Casella}{Levine and
  Casella}{2001}]{levine2001implementations}
Levine, R.~A. and G.~Casella (2001).
\newblock Implementations of the monte carlo em algorithm.
\newblock {\em Journal of Computational and Graphical Statistics\/}~{\em
  10\/}(3), 422--439.

\bibitem[\protect\citeauthoryear{Li, Park, Duraisingham, Strobel, Khan, Soltow,
  Jones, and Pulendran}{Li et~al.}{2013}]{li2013predicting}
Li, S., Y.~Park, S.~Duraisingham, F.~H. Strobel, N.~Khan, Q.~A. Soltow, D.~P.
  Jones, and B.~Pulendran (2013).
\newblock Predicting network activity from high throughput metabolomics.
\newblock {\em PLoS computational biology\/}~{\em 9\/}(7).

\bibitem[\protect\citeauthoryear{Liland}{Liland}{2011}]{liland2011multivariate}
Liland, K.~H. (2011).
\newblock Multivariate methods in metabolomics--from pre-processing to
  dimension reduction and statistical analysis.
\newblock {\em TrAC Trends in Analytical Chemistry\/}~{\em 30\/}(6), 827--841.

\bibitem[\protect\citeauthoryear{Lowsky, Olshansky, Bhattacharya, and
  Goldman}{Lowsky et~al.}{2014}]{lowsky2014heterogeneity}
Lowsky, D.~J., S.~J. Olshansky, J.~Bhattacharya, and D.~P. Goldman (2014).
\newblock Heterogeneity in healthy aging.
\newblock {\em Journals of Gerontology Series A: Biomedical Sciences and
  Medical Sciences\/}~{\em 69\/}(6), 640--649.

\bibitem[\protect\citeauthoryear{Mardia, Kent, and Bibby}{Mardia
  et~al.}{1980}]{Mardia1980}
Mardia, K.~V., J.~T. Kent, and J.~M. Bibby (1980).
\newblock {\em Multivariate analysis}.
\newblock Academic press.

\bibitem[\protect\citeauthoryear{Meinshausen and B{\"u}hlmann}{Meinshausen and
  B{\"u}hlmann}{2006}]{meinshausen2006}
Meinshausen, N. and P.~B{\"u}hlmann (2006).
\newblock High-dimensional graphs and variable selection with the lasso.
\newblock {\em The annals of statistics\/}, 1436--1462.

\bibitem[\protect\citeauthoryear{Nicholson, Lindon, and Holmes}{Nicholson
  et~al.}{1999}]{nicholson1999metabonomics}
Nicholson, J.~K., J.~C. Lindon, and E.~Holmes (1999).
\newblock 'metabonomics': understanding the metabolic responses of living
  systems to pathophysiological stimuli via multivariate statistical analysis
  of biological nmr spectroscopic data.
\newblock {\em Xenobiotica\/}~{\em 29\/}(11), 1181--1189.

\bibitem[\protect\citeauthoryear{Niu and Hoff}{Niu and Hoff}{2014}]{covreg}
Niu, X. and P.~Hoff (2014).
\newblock {\em covreg: A simultaneous regression model for the mean and
  covariance}.
\newblock R package version 1.0.

\bibitem[\protect\citeauthoryear{Niu and Hoff}{Niu and
  Hoff}{2019}]{niu2019joint}
Niu, X. and P.~D. Hoff (2019).
\newblock Joint mean and covariance modeling of multiple health outcome
  measures.
\newblock {\em The annals of applied statistics\/}~{\em 13\/}(1), 321.

\bibitem[\protect\citeauthoryear{Roberts, Souza, Gerszten, and Clish}{Roberts
  et~al.}{2012}]{roberts2012targeted}
Roberts, L.~D., A.~L. Souza, R.~E. Gerszten, and C.~B. Clish (2012).
\newblock Targeted metabolomics.
\newblock {\em Current protocols in molecular biology\/}~{\em 98\/}(1), 30--2.

\bibitem[\protect\citeauthoryear{Robinson, Hyam, Karaman, Pinto, Fiorito, Gao,
  Heard, Jarvelin, Lewis, Pazoki, et~al.}{Robinson
  et~al.}{2018}]{robinson2018determinants}
Robinson, O.~J., M.~C. Hyam, I.~Karaman, R.~C. Pinto, G.~Fiorito, H.~Gao,
  A.~Heard, M.-R. Jarvelin, M.~Lewis, R.~Pazoki, et~al. (2018).
\newblock Determinants of accelerated metabolomic and epigenetic ageing in a uk
  cohort.
\newblock {\em bioRxiv\/}, 411603.

\bibitem[\protect\citeauthoryear{Schott}{Schott}{1991}]{Schott1991}
Schott, J.~R. (1991).
\newblock Some tests for common principal component subspaces in several
  groups.
\newblock {\em Biometrika\/}~{\em 78\/}(4), 771--777.

\bibitem[\protect\citeauthoryear{Stein}{Stein}{1975}]{Stein1975}
Stein, C. (1975).
\newblock Estimation of a covariance matrix.
\newblock {\em Rietz Lecture\/}.

\bibitem[\protect\citeauthoryear{Su and Cook}{Su and
  Cook}{2013}]{su2013estimation}
Su, Z. and R.~D. Cook (2013).
\newblock Estimation of multivariate means with heteroscedastic errors using
  envelope models.
\newblock {\em Statistica Sinica\/}~{\em 23\/}(1), 213--230.

\bibitem[\protect\citeauthoryear{Su, Zhu, Chen, and Yang}{Su
  et~al.}{2016}]{su2016sparse}
Su, Z., G.~Zhu, X.~Chen, and Y.~Yang (2016).
\newblock Sparse envelope model: efficient estimation and response variable
  selection in multivariate linear regression.
\newblock {\em Biometrika\/}~{\em 103\/}(3), 579--593.

\bibitem[\protect\citeauthoryear{Sun and Weckwerth}{Sun and
  Weckwerth}{2012}]{sun2012covain}
Sun, X. and W.~Weckwerth (2012).
\newblock Covain: a toolbox for uni-and multivariate statistics, time-series
  and correlation network analysis and inverse estimation of the differential
  jacobian from metabolomics covariance data.
\newblock {\em Metabolomics\/}~{\em 8\/}(1), 81--93.

\bibitem[\protect\citeauthoryear{Van~der Goot and Nollen}{Van~der Goot and
  Nollen}{2013}]{van2013tryptophan}
Van~der Goot, A.~T. and E.~A. Nollen (2013).
\newblock Tryptophan metabolism: entering the field of aging and age-related
  pathologies.
\newblock {\em Trends in molecular medicine\/}~{\em 19\/}(6), 336--344.

\bibitem[\protect\citeauthoryear{Wang, Zhang, and Li}{Wang
  et~al.}{2019}]{Wang2019}
Wang, W., X.~Zhang, and L.~Li (2019).
\newblock {Common reducing subspace model and network alternation analysis}.
\newblock {\em Biometrics\/}~{\em 75\/}(4), 1109--1120.

\bibitem[\protect\citeauthoryear{Wen and Yin}{Wen and Yin}{2013}]{Wen2013}
Wen, Z. and W.~Yin (2013).
\newblock A feasible method for optimization with orthogonality constraints.
\newblock {\em Mathematical Programming\/}~{\em 142\/}(1-2), 397--434.

\bibitem[\protect\citeauthoryear{Wold}{Wold}{1995}]{wold1995chemometrics}
Wold, S. (1995).
\newblock Chemometrics; what do we mean with it, and what do we want from it?
\newblock {\em Chemometrics and Intelligent Laboratory Systems\/}~{\em
  30\/}(1), 109--115.

\bibitem[\protect\citeauthoryear{Xia, Sinelnikov, Han, and Wishart}{Xia
  et~al.}{2015}]{xia2015metaboanalyst}
Xia, J., I.~V. Sinelnikov, B.~Han, and D.~S. Wishart (2015).
\newblock Metaboanalyst 3.0—making metabolomics more meaningful.
\newblock {\em Nucleic acids research\/}~{\em 43\/}(W1), W251--W257.

\bibitem[\protect\citeauthoryear{Yu and Zhu}{Yu and Zhu}{2010}]{yu2010comments}
Yu, Z. and L.~Zhu (2010).
\newblock Comments on" envelope models for parsimonious and efficient
  multivariate linear regression" by cook, d. and li, b. and chiaromonte.
\newblock {\em Statistica Sinica\/}~{\em 20\/}(3), 988.

\end{thebibliography}

\section*{Supporting Information for ``Reducing Subspace Models for Large-Scale Covariance Regression'' by Alexander Franks}

\subsection*{Maximum Marginal Likelihood Derivations for Previously Proposed Models}

\begin{example_appendix}[\textbf{Response Envelopes, \citet{cook2010envelope}}]
Assume the response envelope model (Equation \ref{eqn:env}). The response envelop model is a special case of model \ref{eqn:shared_subspace} with $\phi_x = X\eta$ with  $\Psi_x = \Psi_1$ independent of $x$.  Further, assume a priori that $\Psi_1 \sim \text{inverse-Wishart}(U_1, \nu_1)$, $\Psi_0 \sim \text{inverse-Wishart}(U_0, \nu_0)$ and that $\eta$ has  the multivariate matrix normal prior $\eta \mid (\Psi_1, V) \sim MN(\eta_0, \Psi_1 \otimes \Lambda_0)$. Then the marginal log-likelihood for $(V, V_\perp)$, after integrating out $\theta = (\Psi_1, \Psi_0, \eta)$ is  
  \begin{equation}
  \label{eqn:marginal_lik}
      \ell(V, V_\perp) \propto -(n + s + \nu_1 + 1 - q)/2\log|V^T(A + U_1)V| - (n-(p-s) + \nu_0 -1)/2\log|V_\perp^T(Y^TY + U_0)V_\perp|
\end{equation}
where 
      $A = (Y - XB_n)^T(Y- X B_n) + B_n^T\Lambda_0B_n$ and $B_n = (X^TX + \Lambda_0)^{-1}X^TY$.
\label{eg:response_envelope_appendix}
\end{example_appendix}

\proof{

Let $Z = YV$, $\beta = \eta V^T$, and $V = [V, V_\perp]$ and $\Sigma = V\Psi_1V^T + V_\perp\Psi_0V^T_\perp$.  The full likelihood can be written as:

\begin{align}
  p(Y \mid V, V_\perp, \eta, \Psi_0, \Psi_1) &\propto |\Sigma|^{-n/2}\etr(-1/2(Y-X\beta)\Sigma^{-1}(Y-X\beta)^{T})\\
  &\propto|V_\perp\Psi_0V_\perp^T + V\Psi_1V^T|^{-n/2}\etr(-1/2(Y-X\beta)(V_\perp\Psi_0^{-1}V_\perp^T + V\Psi_1^{-1}V^T)(Y-X\beta)^{T})\\
&\propto |\Psi_0|^{-n/2}\etr(-1/2V_\perp^T Y^TYV_\perp\Psi_0^{-1}) |\Psi_1|^{-n/2}\etr(-1/2\left[(Z-X\eta)(Z-X\eta)^T\right]\Psi_1^{-1})
\end{align}
\noindent Under the proposed prior distributions the we have
\begin{align}
  &p(\Psi_0, \Psi_1, \eta, Y \mid, X, V, V_\perp) \propto \label{eqn:response_post}\\ 
  & |\Psi_0|^{-n/2}\etr(-1/2\Psi_1^{-1}V_\perp^T Y^TYV_\perp) \times |\Psi_0|^{-(\nu_0 + (p-s)  + 1)/2}\etr(\Psi_0^{-1}V_\perp^TU_0V_\perp) \times\\
 &|\Psi_1|^{-n/2}\etr(-1/2\left[(Z-XH_n)^T(Z-XH_n) + (H_0 - H_n)^T\Lambda_0(H_0-H_n) + (\eta - H_n)^T\Lambda_n(\eta-H_n)\right]\Sigma_1^{-1}) \times\\
  & |\Psi_1|^{-(\nu_1 + s + 1)/2}\etr(\Psi_1^{-1}V^TU_1V) 
\end{align}
\noindent where $H_n = \Lambda_n^{-1}X^TYV$ and $\Lambda_n = (X^TX + \Lambda_0)$.  We compute the marginal likelihood of $V$ and $V_\perp$ as
\begin{align}
  p(Y, \mid V, V_\perp, X) \propto & \int p(V, V_\perp, \Psi_0, \Psi_1, \eta \mid Y, X) d\eta d\Psi_1 d\Psi_0 
\end{align}

\noindent First we isolate all terms in \ref{eqn:response_post} involving $\eta$ and integrate $$\int \etr(\left[\eta - H_n)^T\Lambda_n(\eta-H_n)\right]\Sigma_1^{-1})d\eta \propto |\Psi_1|^{q/2}|$$ since the integrand is an unnnormalized multivariate normal density. Thus
\begin{align}
p(Y, \Psi_0, \Psi_1, \mid V, V_\perp, X) \propto& |\Psi_0|^{-n/2}\etr(-1/2V_\perp\Psi_0^{-1}V_\perp^T Y^TY) \times\\
  & |\Psi_1|^{-(n-q)  /2}\etr(-1/2\left[(Z-XH_n)^T(Z-XH_n) + (H_0 - H_n)^T\Lambda_0(H_0-H_n)\right]\Psi_1^{-1}) \times\\
  & |\Psi_0|^{-(\nu_0 + (p-s)  + 1)/2}\etr(\Psi_0^{-1}V_\perp^T U_0V_\perp) |\Psi_1|^{-(\nu_1 + s + 1)/2}\etr(\Psi_1^{-1}V^TU_1V) \\
\end{align}

\noindent Next we isolate terms involving $\Psi_1$ and integrate:
\begin{align}
  \int    |\Psi_1|^{-(n-q + \nu_1)/2}\etr(-1/2\left[(Z-XH_n)^T(Z-XH_n) + (H_0 - H_n)^T\Lambda_0(H_0-H_n)  + U_1\right]\Psi_1^{-1}) d\Psi_1
\end{align}
\noindent which is proportional to an inverse-Wishart($(Z-XH_n)^T(Z-XH_n) + (H_0 - H_n)^T\Lambda_0(H_0-H_n) + U_1, (n-q + \nu_1)$) density with normalizing constant proportional to
\begin{align}
&|(Z-XH_n)^T(Z-XH_n) + (H_0 - H_n)^T\Lambda_0(H_0-H_n) + U_1|^{-(n-q + s + \nu_1)/2} = \\
&|V^T(Y-XB_n)^T(Y-XB_n) + (B_0 - B_n)^T\Lambda_0(B_0-B_n) + U_1)V|^{-(n-q + s + \nu_1)/2} = \\
&|V^TAV|^{-(n-q + s + \nu_1)/2} \label{eqn:psi1int}
\end{align}

\noindent Finally we integrate out $\Sigma_0$, 
\begin{align}
  &= \int |\Sigma_0|^{-n/2}\etr(-1/2(V_\perp^T Y^TYV_\perp + V_\perp^TU_0V_\perp)\Sigma_0^{-1}) d\Sigma_0 \\
  &= |V_\perp^T(Y^TY + U_0)V_\perp|^{-\frac{(n + (p-s) + \nu_0 + 1)}{2}} \label{eqn:psi0int}
\end{align}
\noindent Combining equations \ref{eqn:psi1int} and \ref{eqn:psi0int} we have the marginal liklihood of $(V, V_\perp)$:
\begin{align*}
      p(Y \mid V, V_\perp, X) &\propto |VAV^T|^{-(n + s + \nu_1 + 1 - q)/2} |V^T(Y^TY + U_0)V|^{- (n + s + \nu_0 + 1)/2} \\
  A &= (Y^T - B_n^T)(Y- X B_n) + B_n^T\Lambda_0B_n + U_1\\
  B_n &= (X^TX + \Lambda_0)^{-1}X^TY\\
\end{align*}

}

With the non-informative improper prior distributions $U_i = \mathbf{0}$ and $\Lambda_0 = \mathbf{0}$ and $n \gg p$, this objective function is approximately proportional to the  objective function originally proposed by \citet{cook2010envelope}: $\ell(V, V_\perp) = \log(|V^TAV|) + \log(V_\perp^TY^TYV_\perp)$ .  In small $n$ settings, the difference in between these two objectives may be nontrivial.  It should also be noted Equation \ref{eqn:marginal_lik} can be expressed as a function of $V$ only, since the subspace spanned by $V_\perp$ is identified by $V$ alone \footnote{\citet{cook2013envelopes} use the modified objective $\ell(V) \propto \log|V^T(A )V| + \log|V^T(Y^TY )^{-1}V|$, which asymptotically has the same minimizer as Equation \ref{eqn:marginal_lik} when the envelope model holds.}.  The maximum marginal likelihood perspective also provides a principled way of including additional regularization by specifying appropriate  prior parameters \citep[see e.g.][]{yu2010comments}. \citet{su2016sparse} proposed an extension of to the envelope model to high-dimensional settings by a assuming a sparse envelope and modifying the usual envelope objective is augmented with a group lasso penalty on $V$.  In our framework, such a penalty would equivalently be viewed as prior distribution over the space of semi-orthogonal matrices, $V$.


\begin{example_appendix}[\textbf{Shared Subspace Covariance Models, \citet{franks2019shared}}]
Assume the shared subspace model proposed by \citet{franks2019shared}, that is $\phi_x = 0$ and $x$ is a single categorical predictor.  Let $Y_k$ be the $n_k \times p$ matrix of observations for which $x$ has level $k$ and let $\Psi_x = \Psi_k$ denote the corresponding covariance matrix of $Y_k$.  They assume a spiked covariance model, for which the data projected onto the subspace orthogonal to $V$ is isotropic, i.e. for which $\Psi_0 = \sigma^2I$.  $\Psi_k \sim inverse-Wishart(U_k, \nu_k)$ and $\sigma^2 \sim \text{inverse-Gamma}(\alpha, \kappa)$. Then the marginal log-likelihood, after integrating out $\theta = (\Psi_1, ... \Psi_k, \sigma^2)$ is

The complete likelihood can be expressed as:

\begin{align*}
  \ell(V)  & \propto \sum_{k=1}^K  \left[-(n_k - \nu_1)/2\log |V^T (Y_k^TY_k + U_k) V| - (n_k(p - s) + \alpha)\log(\rm{tr}[(I - V V^T)Y_k^TY_k]/2 + \kappa)\right]
\end{align*}
\end{example_appendix}

\proof{ 
\begin{align}
  p(Y_1, \ldots Y_k \mid V, V_\perp \sigma^2_1 \ldots \sigma^2_k, \Psi_1) &\propto \prod |\Sigma_k|^{-n_1/2}\etr(-1/2Y_kY_k^T\Sigma_k^{-1})\\
  &\propto \prod |V\Psi_kV^T + \sigma^2_k(I - VV^T)|^{-n_1/2}\etr(-1/2Y_kY_k^T(V\Psi_kV^T + \sigma^2_k)^{-1})\\
&\propto (\sigma_k^2)^{-n_1(p-s)}\etr(\frac{1}{\sigma_k^2}(I-VV^T)Y_k^TY_k)|\Psi_k|^{-n_1/2}\etr(-1/2Y_kY_k^TV\Psi_k^{-1}V^T) 
\end{align}

\noindent We consider the marginal likelihood for an arbitrary group $k$, as each group is independent conditional on $V$.  First, we integrate $\sigma_k^2$ given that $\sigma_k^2 \sim \text{inverse-Gamma}(\alpha, \kappa)$:

\begin{align}
  p(Y_k \mid V, \Psi_k) &\propto \int p(Y_k \mid V, \Psi_k, \sigma_k^2)p(\sigma_2k^2) d \sigma_k^2 \\
&\propto \left( \int (\sigma_k^2)^{-(p-s)}\etr(\frac{1}{\sigma_k^2}(I-VV^T)Y_k^TY_k)  (\sigma^2_k)^{\alpha-1}\etr(\kappa/\sigma^2_k)d\sigma_k^2\right)|\Psi_k|^{-n_k/2}\etr(-1/2Y_k^TY_k^TV\Psi_k^{-1}V^T) 
\end{align}
\noindent The integrand is a unnormalized inverse-Gamma($(n_k(p - s) + \alpha), \rm{tr}[(I - V V^T)Y_k^TY_k]/2 + \kappa))$ density with normalizing constant proportional to $\rm{tr}[(I - V V^T)Y_k^TY_k]/2 + \kappa)^{- (n_k(p - s) + \alpha)}$, so that

\begin{align}
    p(Y_k \mid V, \Psi_k) \propto \rm{tr}[(I - V V^T)Y_k^TY_k]/2 + \kappa)^{- (n_1p - s) + \alpha)}|\Psi_k|^{-n_k/2}\etr(-1/2V^TY_k^TY_k^TV\Psi_k^{-1}) 
\end{align}

\noindent Next, we integrate out $\Psi_k$ given that $\Psi_k \sim \text{inverse-Wishart}(U_k, \nu_k)$.

\begin{align}
  p(Y_k \mid V) &\propto \int p(Y_k \mid V, \Psi_k)p(\Psi_k^2) d \Psi_k \\
&\propto \int \left(|\Psi_k|^{-n_k/2}\etr(-1/2Y_k^TY_k^TV\Psi_k^{-1}V^T)  |\Psi_k|^{-(\nu_k+s+1)/2} \etr(-1/2U_k\Psi^{-1})d\Psi_1 \right)
\end{align}

\noindent The integrand is a unnormalized inverse-Wishart($Y_k^TY_k + U_k, n_k + \nu_1))$ density with normalizing constant proportional to $|V^T (Y_k^TY_k + U_k) V|^{-(n_k/2 - \nu_k)/2}$ so that
\begin{align}
 p(Y_k \mid V) &\propto |V^T (Y_k^TY_k + U_k) V|^{-(n_k - \nu_k)/2}\rm{tr}[(I - V V^T)Y_k^TY_k]/2 + \kappa)^{- (n_k(p - s) + \alpha)}
\end{align}

\noindent The objective function is then available as $\text{log} \prod_k p(Y_k \mid V)$ using the above result.

}

\noindent \citet{franks2019shared} propose a method for inferring the shared subspace using the EM algorithm, although the marginal distribution can be derived analytically, as above.

\subsection*{Proof of Theorem 1}
\begin{theorem_appendix}
Assume a model satisfying Equation \ref{eqn:shared_subspace}.  Let the $\Psi_0 \sim \text{inverse-Wishart}(U_0, \nu_0)$, and assume the $\Psi_{x_i}$ and mean $\phi_{x_i}$ are known.  The marginal log-likelihood for $[V~ V_\perp]$, integrating over, $\Psi_0$, with $\Psi_{x_i}$ and $\phi_{x_i}$ assumed known is:
\begin{align}
\ell(V, V_\perp; Y, X, \Psi_x, \phi_x) &=  
    -\frac{1}{2}\sum_i\left[ y_i V \Psi_{x_i}^{-1}V^Ty_i^T - 2\phi_{x_i}\Psi_{x_i}^{-1}V^Ty_i^T\right]\\
\nonumber &~~~ - \frac{(n+(p-s)+\nu_0+1)}{2}\log|V_\perp^T(Y^TY + U_0)V_\perp|
\end{align}
\end{theorem_appendix}

\proof{ 
\begin{align}
  p(Y \mid X, V, V_\perp \Psi_x, \Psi_0, \phi_x) &\propto \prod_i |\Sigma_{x_i}|^{-1/2}\etr(-1/2(y_i - \phi_{x_i}V)(y_i-\phi_{x_i}V)^T\Sigma_{x_i}^{-1}) \\
  &\propto \prod_i |V\Psi_{x_i}V^T + V_\perp\Psi_0V_\perp^T|^{-1/2}\etr(-1/2(Y_i - \phi_{x_i}V)(Y_i-\phi_{x_i}V)^T(V\Psi_{x_i}^{-1}V^T + V_\perp\Psi_0^{-1}V_\perp^T)) \\
&\propto \left(\prod_i |\Psi_{x_i}|^{-1/2}\etr(-1/2(Y_i - \phi_{x_i}V)(Y_i-\phi_{x_i}V)^TV\Psi_{x_i}^{-1}V^T)\right) \\ & \times |\Psi_0|^{-n/2}\etr(-1/2YY^TV_\perp\Psi_0^{-1}V_\perp^T))  \\
&\propto \left(\prod_i |V\Psi_{x_i}V^T|^{-1/2}\etr(-1/2(Y_i^TY_iV\Psi^{-1}_{x_i} - 2\phi_{x_i}\Psi^{-1}_{x_i}V)\right)\\ & \times |V_\perp\Psi_0V_\perp^T|^{-n/2}\etr(-1/2YY^TV_\perp\Psi_0^{-1}V_\perp^T))
\end{align}

\noindent Under the inverse-Wishart prior we integrate out $\Psi_0$:

\begin{align}
  p(Y \mid X, V, V_\perp \Psi_x,\phi_x) &\propto   \int p(Y \mid X, V, V_\perp \Psi_x,\phi_x)p(\Psi_0) d\Psi_0 \\
  &\propto \left(\prod_i |V\Psi_{x_i}V^T|^{-1/2}\etr(-1/2(y_iy_iV\Psi^{-1}_{x_i} - 2\phi_{x_i}\Psi^{-1}_{x_i}V)\right)\\ & \times \int |V_\perp\Psi_0V_\perp^T|^{-n/2}\etr(-1/2YY^TV_\perp\Psi_0^{-1}V_\perp^T)) |V_\perp\Psi_0V_\perp^T|^{-(\nu_0 -1)}\etr(V_\perp\Psi_0^{-1}V_\perp^TU_0) d\Psi_0
\end{align}

\noindent The integrand is a unnormalized inverse-Wishart($V_\perp (Y^TY + U_0)V_\perp, n + \nu_0))$ density with normalizing constant which is proportional to $|V_\perp^T (Y^TY + U_0)V_\perp V|^{-(n - \nu_0)/2}$ so that
\begin{align}
p(Y \mid X, V, V_\perp \Psi_x,\phi_x) &\propto     &\propto \left(\prod_i |V\Psi_{x_i}V^T|^{-1/2}\etr(-1/2(y_i^Ty_iV\Psi^{-1}_{x_i} - 2\phi_{x_i}\Psi^{-1}_{x_i}V)\right)|V_\perp^T (Y^TY + U_0) V_\perp|^{-(n + (p-s) + \nu_0)/2}
\end{align}

}

\begin{corollary}

Assume a model satisfying Equation \ref{eqn:spiked_model} and a prior for $\sigma^2 \sim \text{inverse-Gamma}(\alpha, \kappa)$.  The marginal log-likelihood for $V$, integrating over $\sigma^2$, assuming $\Psi_{x_i}$ and $\phi_{x_i}$ are known is:
\begin{align}
\ell(V; Y, X, \Psi_x, \phi_x) &=  
    -\frac{1}{2}\sum_i\left[ y_i V \Psi_{x_i}^{-1}V^Ty_i^T - 2\phi_{x_i}\Psi_{x_i}^{-1}V^Ty_i^T\right]\\
\nonumber &~~~ - \left(\frac{(n(p-s)}{2} - \alpha\right)\log(\frac{1}{2}||Y||_F^2 - \frac{1}{2}||YV||_F^2 + \kappa)
\end{align}
\end{corollary}

\proof{ 

\begin{align}
  p(Y \mid X, V, V_\perp \Psi_x, \Psi_0, \phi_x) &\propto \prod_i |\Sigma_{x_i}|^{-1/2}\etr(-1/2(Y_i - \phi_{x_i}V)(Y_i-\phi_{x_i}V)^T\Sigma_{x_i}^{-1}) \\
  &\propto \prod_i |V\Psi_{x_i}V^T + \sigma^2(I-VV^T)|^{-1/2}\etr(-1/2(Y_i - \phi_{x_i}V)(Y_i-\phi_{x_i}V)^T(V\Psi_{x_i}^{-1}V^T + \frac{1}{\sigma^2}(I-VV^T)) \\
&\propto \prod_i |\Psi_{x_i}|^{-1/2}\etr(-1/2(Y_i - \phi_{x_i}V)(Y_i-\phi_{x_i}V)^TV\Psi_{x_i}^{-1}V^T) \\ & 
\times (\sigma^2)^{-n(p-s)/2}\etr(-1/2YY^T(I - VV^T)/\sigma^2)  \\
\end{align}

Under the inverse-Gamma prior, we can integrate out $\sigma^2$

\begin{align}
  p(Y \mid X, V, \Psi_x,\phi_x) &\propto   \int p(Y \mid X, V, \Psi_x,\phi_x)p(\sigma^2) d\Psi_0 \\
  &\propto \prod_i |V\Psi_{x_i}V^T|^{-1/2}\etr(-1/2(Y_i^TY_iV\Psi^{-1}_{x_i} - 2\phi_{x_i}\Psi^{-1}_{x_i}V)\\ & \times \int (\sigma^2)^{-n(p-s)/2}\etr(-1/2YY^T(I-VV^T)/\sigma^2)) (\sigma^2)^{-\alpha}e^{(\kappa/\sigma^2)}d\sigma^2
\end{align}

\noindent The integrand is a unnormalized inverse-Gamma($, n(p-s) + \alpha)$ density with normalizing constant which is proportional to 

\begin{align} 
|\text{tr}(-1/2YY^T(I-VV^T)) + \kappa|^{-(n(p-s)/2 + \alpha)} =  (1/2||Y||_F^2 - 1/2||YV||_F^2 + \kappa)^{-(n(p-s)/2 + \alpha)}
\end{align}
so that
\begin{align}
&p(Y \mid X, V, V_\perp \Psi_x,\phi_x) \propto \\
&\left(\prod_i |V\Psi_{x_i}V^T|^{-1/2}\etr(-1/2(Y_i^TY_iV\Psi^{-1}_{x_i} - 2\phi_{x_i}\Psi^{-1}_{x_i}V)\right)(1/2||Y||_F^2 - 1/2||YV||_F^2 + \kappa)^{-(n(p-s)/2 + \alpha)}
\end{align}

}

\pagebreak

\subsection*{Metabolomic Analysis of Mean-level Results}

\begin{table}[h!]
\begin{center}
\begin{tabular}{l|r|r|r}
\textbf{Metabolite} & \textbf{P-value} & \textbf{T-statistic} & \textbf{Q-value}\\
\hline
 HIAA & 0.000 & 6.189341 & 0.0000000\\
\hline
 Cystine & 0.000 & 5.893680 & 0.0000000\\
\hline
 Kynurenine & 0.000 & 5.806419 & 0.0000000\\
\hline
 4-Aminobutyric acid & 0.000 & -5.177550 & 0.0000000\\
\hline
 Glycerol 3-phosphate & 0.000 & -4.444250 & 0.0000000\\
\hline
 Adenosine & 0.000 & -4.345212 & 0.0000000\\
\hline
 Uridine & 0.000 & -4.173488 & 0.0000000\\
\hline
 Decanoylcarnitine & 0.000 & 3.977997 & 0.0000000\\
\hline
 Acetamide & 0.000 & -3.911912 & 0.0000000\\
\hline
 Uracil & 0.000 & -3.441107 & 0.0000000\\
\hline
 Aspartic acid & 0.000 & 3.411236 & 0.0000000\\
\hline
 Acetylglucosamine & 0.000 & 3.390308 & 0.0000000\\
\hline
 Xanthine & 0.000 & 3.360388 & 0.0000000\\
\hline
 Carnitine & 0.000 & 3.220702 & 0.0000000\\
\hline
 alpha-ketoisovaleric acid & 0.000 & -3.032598 & 0.0000000\\
\hline
 Acetylglycine & 0.000 & 2.880037 & 0.0000000\\
\hline
 Glycylproline & 0.002 & 3.288615 & 0.0127059\\
\hline
 Glycine & 0.006 & 2.790219 & 0.0360000\\
\hline
 Levulinic acid & 0.006 & -2.746456 & 0.0341053\\
\hline
 4-Methylvaleric acid & 0.006 & -2.647026 & 0.0324000\\
\hline
 Adenosyl-L-homocysteine & 0.008 & 2.960800 & 0.0411429\\
\hline
 Indole-3-acetic acid & 0.008 & 2.765461 & 0.0392727\\
\hline
 Glycoursodeoxycholic acid & 0.010 & -2.618862 & 0.0469565\\
\hline
 Glycohyodeoxycholic acid & 0.010 & -2.604362 & 0.0450000\\
\hline
 Fructose & 0.010 & -2.465946 & 0.0432000\\
\hline
 Indole & 0.012 & -2.564810 & 0.0498462\\
\hline
\end{tabular}
\caption{Targeted data.  Metabolites whose mean levels vary significantly with age.}
\end{center}
\end{table}

\begin{table}[h!]
\begin{center}
\begin{tabular}{l|r}
\textbf{Pathway} & \textbf{P-value}\\
\hline
Vitamin B3 (nicotinate and nicotinamide) metabolism & 0.0105033\\
\hline
Nitrogen metabolism & 0.0119318\\
\hline
Phosphatidylinositol phosphate metabolism & 0.0125200\\
\hline
Drug metabolism - cytochrome P450 & 0.0184018\\
\hline
Vitamin E metabolism & 0.0275607\\
\hline
Glutamate metabolism & 0.0275607\\
\hline
\end{tabular}
\caption{Untargeted data. Pathways associated with metabolites whose mean levels vary with age. }
\label{table:1}
\end{center}
\end{table}

\begin{table}[h!]
\begin{center}
\begin{tabular}{l|r|r|r}
\textbf{Metabolite} & \textbf{P-value} & \textbf{T-statistic} & \textbf{Q-value}\\
\hline
Stearic acid & 0 & -3.319559 & 0\\
\hline
N,N'-Dicyclohexylurea & 0 & 2.770768 & 0\\
\hline
\end{tabular}
\caption{Targeted data.  Metabolites whose mean levels vary significantly with sex.}
\end{center}
\end{table}

\pagebreak
\subsection*{Select Covariance Regression Results - Sex}

\begin{figure}
     \centering
     \begin{subfigure}
    \centering
    \includegraphics[width=0.75\textwidth]{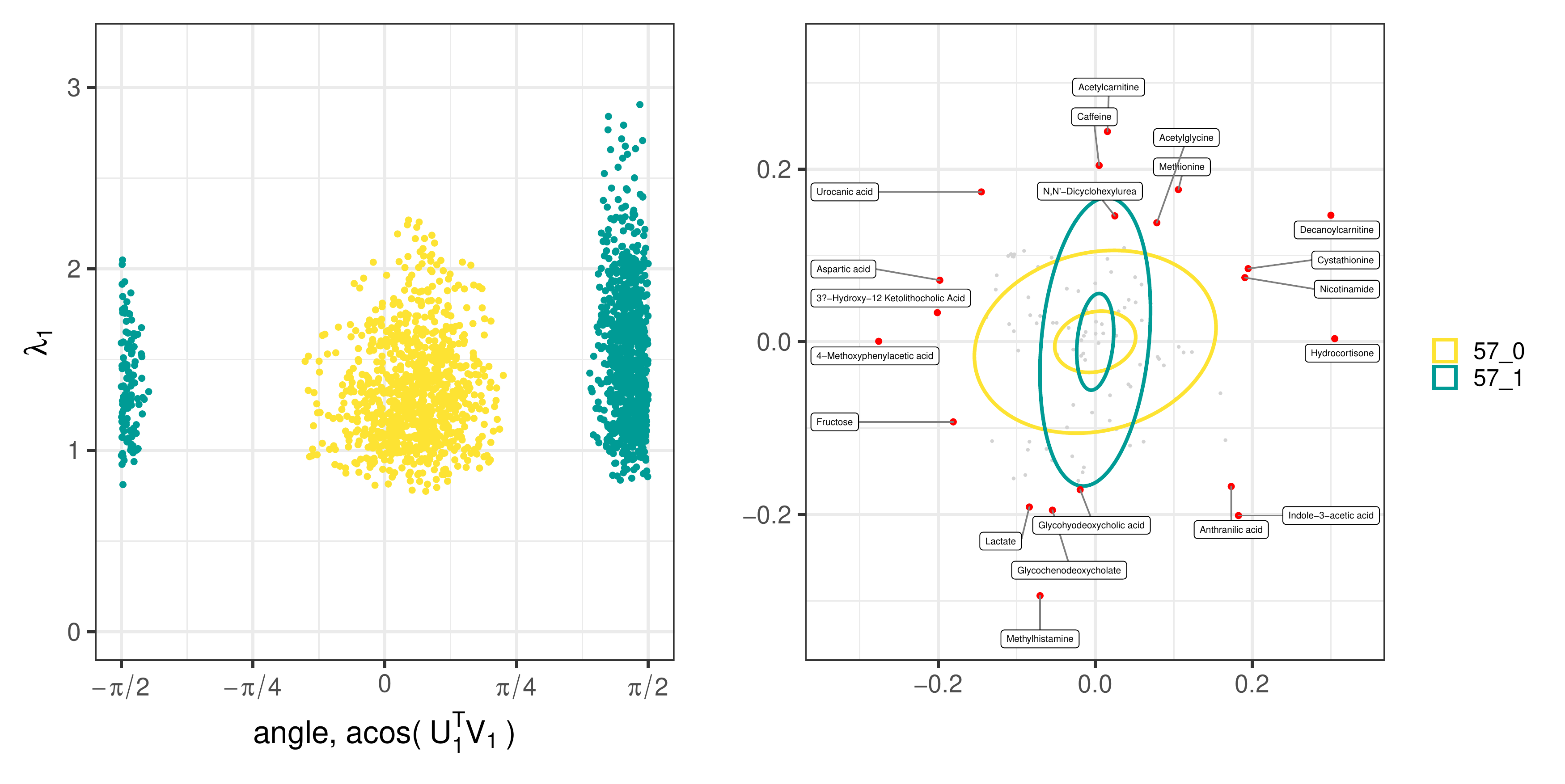}
\label{fig:aging_targeted}
     \end{subfigure} 
     \hfill \\
     \begin{subfigure}
         \centering
    \includegraphics[width=0.75\textwidth]{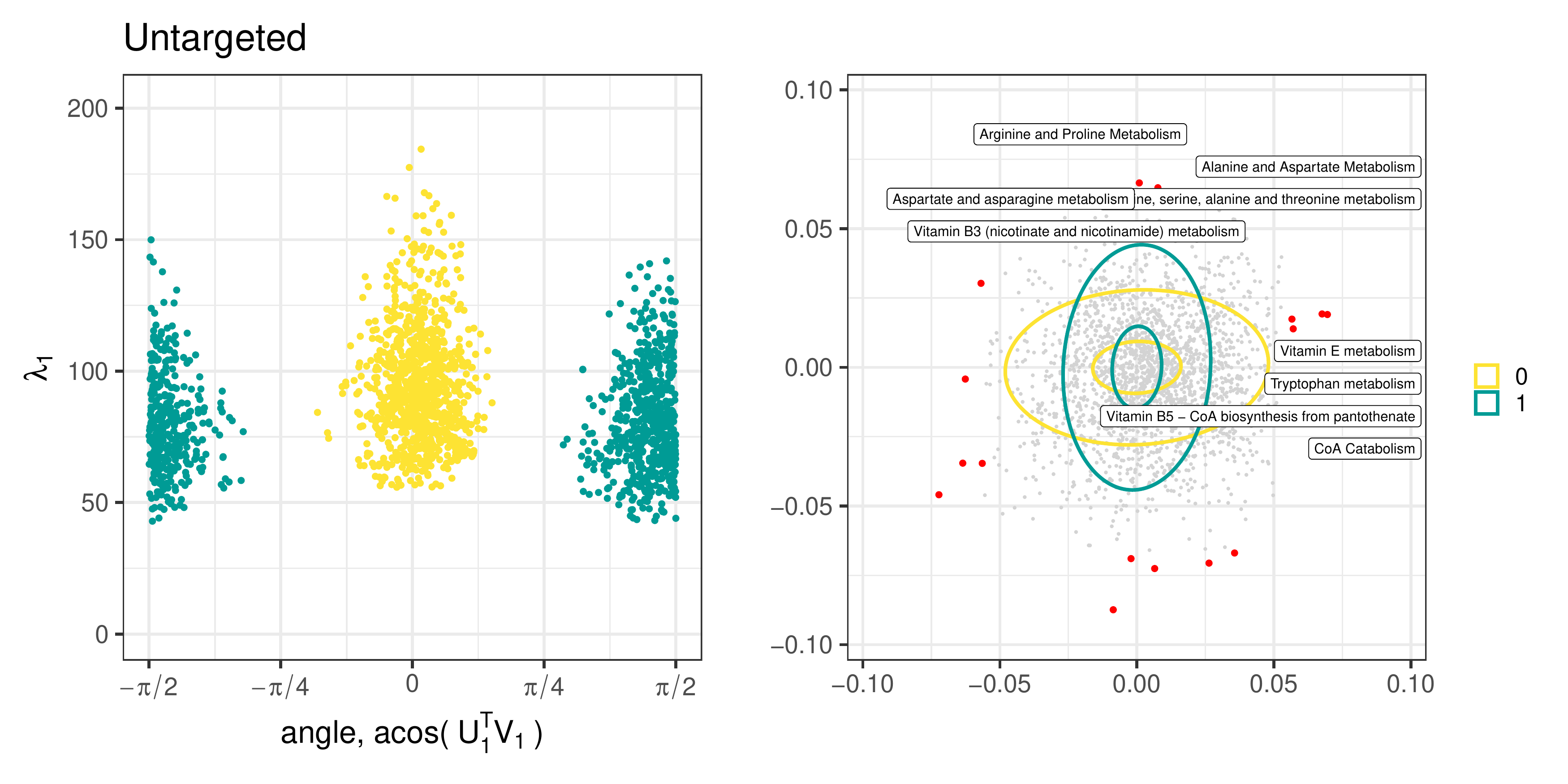}
\label{fig:aging_untargeted}
     \end{subfigure}
         \caption{Posterior summaries for sex-dependent covariance matrices on the targeted (top) and untargeted (bottom) dataset.  Following \citet{franks2019shared}, we summarize the posterior distribution of the covariance matrices in terms of their eigenvalue and eigenvectors. Left) posterior samples of the largest eigenvalue and orientation of the first principal component on a two-dimensional subspace of $VV^T$ that describes the largest difference between the youngest and oldest male individuals. Right) A variant of the PCA biplot on the subspace of variation.  Contours depict the posterior mean covariance matrices at different ages.  Points reflect the metabolites with the largest factor loadings.  For targeted data we include the names of metabolites with the largest factor loadings, whereas for the untargeted analysis, we include pathways enrichment analysis \texttt{mummichog}.}
        \label{fig:results_appendix}
\end{figure}

\end{document}